\documentclass[longauth]{aa} 

\usepackage{graphicx}
\usepackage{txfonts}

\usepackage[hidelinks,colorlinks=true,linkcolor=blue,citecolor=blue,backref=page]{hyperref}
\usepackage{color}
\usepackage{textcomp,gensymb}
\usepackage{amsmath}
\usepackage{rotating}
\usepackage{caption}
\usepackage{comment}
\usepackage{placeins}

\makeatletter
\renewcommand*\aa@pageof{, page \thepage{} of \pageref*{LastPage}}
\makeatother

\newcommand{\spitzer}{\texttt{Spitzer}\xspace}
\newcommand{\stname}{WASP-12\xspace}
\newcommand{\plname}{WASP-12\,b\xspace}

\newcommand{\cheops}{\texttt{CHEOPS}\xspace}
\newcommand{\hst}{\texttt{HST}\xspace}
\newcommand{\tess}{\texttt{TESS}\xspace}

\newcommand{\hf}{\ensuremath{h_2}\xspace}
\newcommand{\Rv}{\ensuremath{R_{\mathrm{v}}}\xspace}

\newcommand\pyratbay{\textsc{Pyrat Bay}}

\usepackage[normalem]{ulem}

\begin{document}

\title{The tidal deformation and atmosphere of WASP-12\,b from its phase curve\thanks{Based on data from CHEOPS guaranteed time observations (GTO) with Program IDs: CH\_PR100013, CH\_PR100016, and CH\_PR330093}, \thanks{The CHEOPS photometric time-series data used
in this paper are available in electronic form at the CDS via anonymous ftp to cdsarc.u-strasbg.fr (130.79.128.5) or via http://cdsweb.u-
strasbg.fr/cgi-bin/qcat?J/A+A/}}
\titlerunning{The tidal deformation and atmosphere of WASP-12b}

\author{
B. Akinsanmi\inst{1} $^{\href{https://orcid.org/0000-0001-6519-1598}{\includegraphics[scale=0.5]{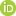}}}$, 
S. C. C. Barros\inst{2,3} $^{\href{https://orcid.org/0000-0003-2434-3625}{\includegraphics[scale=0.5]{figures/orcid.jpg}}}$, 
M. Lendl\inst{1} $^{\href{https://orcid.org/0000-0001-9699-1459}{\includegraphics[scale=0.5]{figures/orcid.jpg}}}$, 
L. Carone\inst{4}, 
P. E. Cubillos\inst{5,4}, 
A. Bekkelien\inst{1}, 
A. Fortier\inst{6,7} $^{\href{https://orcid.org/0000-0001-8450-3374}{\includegraphics[scale=0.5]{figures/orcid.jpg}}}$, 
H.-G. Florén\inst{8}, 
A. Collier Cameron\inst{9} $^{\href{https://orcid.org/0000-0002-8863-7828}{\includegraphics[scale=0.5]{figures/orcid.jpg}}}$, 
G. Boué\inst{10} $^{\href{https://orcid.org/0000-0002-5057-7743}{\includegraphics[scale=0.5]{figures/orcid.jpg}}}$, 
G. Bruno\inst{11} $^{\href{https://orcid.org/0000-0002-3288-0802}{\includegraphics[scale=0.5]{figures/orcid.jpg}}}$, 
B.-O. Demory\inst{7,6} $^{\href{https://orcid.org/0000-0002-9355-5165}{\includegraphics[scale=0.5]{figures/orcid.jpg}}}$, 
A. Brandeker\inst{8} $^{\href{https://orcid.org/0000-0002-7201-7536}{\includegraphics[scale=0.5]{figures/orcid.jpg}}}$, 
S. G. Sousa\inst{2} $^{\href{https://orcid.org/0000-0001-9047-2965}{\includegraphics[scale=0.5]{figures/orcid.jpg}}}$, 
T. G. Wilson\inst{12} $^{\href{https://orcid.org/0000-0001-8749-1962}{\includegraphics[scale=0.5]{figures/orcid.jpg}}}$, 
A. Deline\inst{1}, 
A. Bonfanti\inst{4} $^{\href{https://orcid.org/0000-0002-1916-5935}{\includegraphics[scale=0.5]{figures/orcid.jpg}}}$, 
G. Scandariato\inst{11} $^{\href{https://orcid.org/0000-0003-2029-0626}{\includegraphics[scale=0.5]{figures/orcid.jpg}}}$, 
M. J. Hooton\inst{13} $^{\href{https://orcid.org/0000-0003-0030-332X}{\includegraphics[scale=0.5]{figures/orcid.jpg}}}$, 
A. C. M. Correia\inst{48} $^{\href{https://orcid.org/0000-0002-8946-8579}{\includegraphics[scale=0.5]{figures/orcid.jpg}}}$, 
O. D. S. Demangeon\inst{2,3} $^{\href{https://orcid.org/0000-0001-7918-0355}{\includegraphics[scale=0.5]{figures/orcid.jpg}}}$, 
A. M. S. Smith\inst{14} $^{\href{https://orcid.org/0000-0002-2386-4341}{\includegraphics[scale=0.5]{figures/orcid.jpg}}}$, 
V. Singh\inst{11} $^{\href{https://orcid.org/0000-0002-7485-6309}{\includegraphics[scale=0.5]{figures/orcid.jpg}}}$, 
Y. Alibert\inst{7,6} $^{\href{https://orcid.org/0000-0002-4644-8818}{\includegraphics[scale=0.5]{figures/orcid.jpg}}}$, 
R. Alonso\inst{15,16} $^{\href{https://orcid.org/0000-0001-8462-8126}{\includegraphics[scale=0.5]{figures/orcid.jpg}}}$, 
J. Asquier\inst{17}, 
T. Bárczy\inst{18} $^{\href{https://orcid.org/0000-0002-7822-4413}{\includegraphics[scale=0.5]{figures/orcid.jpg}}}$, 
D. Barrado Navascues\inst{19} $^{\href{https://orcid.org/0000-0002-5971-9242}{\includegraphics[scale=0.5]{figures/orcid.jpg}}}$, 
W. Baumjohann\inst{4} $^{\href{https://orcid.org/0000-0001-6271-0110}{\includegraphics[scale=0.5]{figures/orcid.jpg}}}$, 
M. Beck\inst{1} $^{\href{https://orcid.org/0000-0003-3926-0275}{\includegraphics[scale=0.5]{figures/orcid.jpg}}}$, 
T. Beck\inst{6}, 
W. Benz\inst{6,7} $^{\href{https://orcid.org/0000-0001-7896-6479}{\includegraphics[scale=0.5]{figures/orcid.jpg}}}$, 
N. Billot\inst{1} $^{\href{https://orcid.org/0000-0003-3429-3836}{\includegraphics[scale=0.5]{figures/orcid.jpg}}}$, 
X. Bonfils\inst{20} $^{\href{https://orcid.org/0000-0001-9003-8894}{\includegraphics[scale=0.5]{figures/orcid.jpg}}}$, 
L. Borsato\inst{21} $^{\href{https://orcid.org/0000-0003-0066-9268}{\includegraphics[scale=0.5]{figures/orcid.jpg}}}$, 
C. Broeg\inst{6,7} $^{\href{https://orcid.org/0000-0001-5132-2614}{\includegraphics[scale=0.5]{figures/orcid.jpg}}}$, 
M. Buder\inst{22}, 
S. Charnoz\inst{23} $^{\href{https://orcid.org/0000-0002-7442-491X}{\includegraphics[scale=0.5]{figures/orcid.jpg}}}$, 
Sz. Csizmadia\inst{14} $^{\href{https://orcid.org/0000-0001-6803-9698}{\includegraphics[scale=0.5]{figures/orcid.jpg}}}$, 
M. B. Davies\inst{24} $^{\href{https://orcid.org/0000-0001-6080-1190}{\includegraphics[scale=0.5]{figures/orcid.jpg}}}$, 
M. Deleuil\inst{25} $^{\href{https://orcid.org/0000-0001-6036-0225}{\includegraphics[scale=0.5]{figures/orcid.jpg}}}$, 
L. Delrez\inst{26,27} $^{\href{https://orcid.org/0000-0001-6108-4808}{\includegraphics[scale=0.5]{figures/orcid.jpg}}}$, 
D. Ehrenreich\inst{1,28} $^{\href{https://orcid.org/0000-0001-9704-5405}{\includegraphics[scale=0.5]{figures/orcid.jpg}}}$, 
A. Erikson\inst{14}, 
J. Farinato\inst{21}, 
L. Fossati\inst{4} $^{\href{https://orcid.org/0000-0003-4426-9530}{\includegraphics[scale=0.5]{figures/orcid.jpg}}}$, 
M. Fridlund\inst{29,30} $^{\href{https://orcid.org/0000-0002-0855-8426}{\includegraphics[scale=0.5]{figures/orcid.jpg}}}$, 
D. Gandolfi\inst{31} $^{\href{https://orcid.org/0000-0001-8627-9628}{\includegraphics[scale=0.5]{figures/orcid.jpg}}}$, 
M. Gillon\inst{26} $^{\href{https://orcid.org/0000-0003-1462-7739}{\includegraphics[scale=0.5]{figures/orcid.jpg}}}$, 
M. Güdel\inst{32}, 
M. N. Günther\inst{17} $^{\href{https://orcid.org/0000-0002-3164-9086}{\includegraphics[scale=0.5]{figures/orcid.jpg}}}$, 
A. Heitzmann\inst{1}, 
Ch. Helling\inst{4,33}, 
S. Hoyer\inst{25} $^{\href{https://orcid.org/0000-0003-3477-2466}{\includegraphics[scale=0.5]{figures/orcid.jpg}}}$, 
K. G. Isaak\inst{17} $^{\href{https://orcid.org/0000-0001-8585-1717}{\includegraphics[scale=0.5]{figures/orcid.jpg}}}$, 
L. L. Kiss\inst{34,35}, 
K. W. F. Lam\inst{14} $^{\href{https://orcid.org/0000-0002-9910-6088}{\includegraphics[scale=0.5]{figures/orcid.jpg}}}$, 
J. Laskar\inst{10} $^{\href{https://orcid.org/0000-0003-2634-789X}{\includegraphics[scale=0.5]{figures/orcid.jpg}}}$, 
A. Lecavelier des Etangs\inst{36} $^{\href{https://orcid.org/0000-0002-5637-5253}{\includegraphics[scale=0.5]{figures/orcid.jpg}}}$, 
D. Magrin\inst{21} $^{\href{https://orcid.org/0000-0003-0312-313X}{\includegraphics[scale=0.5]{figures/orcid.jpg}}}$, 
P. F. L. Maxted\inst{37} $^{\href{https://orcid.org/0000-0003-3794-1317}{\includegraphics[scale=0.5]{figures/orcid.jpg}}}$, 
M. Mecina\inst{32}, 
C. Mordasini\inst{6,7}, 
V. Nascimbeni\inst{21} $^{\href{https://orcid.org/0000-0001-9770-1214}{\includegraphics[scale=0.5]{figures/orcid.jpg}}}$, 
G. Olofsson\inst{8} $^{\href{https://orcid.org/0000-0003-3747-7120}{\includegraphics[scale=0.5]{figures/orcid.jpg}}}$, 
R. Ottensamer\inst{32}, 
I. Pagano\inst{11} $^{\href{https://orcid.org/0000-0001-9573-4928}{\includegraphics[scale=0.5]{figures/orcid.jpg}}}$, 
E. Pallé\inst{15,16} $^{\href{https://orcid.org/0000-0003-0987-1593}{\includegraphics[scale=0.5]{figures/orcid.jpg}}}$, 
G. Peter\inst{22} $^{\href{https://orcid.org/0000-0001-6101-2513}{\includegraphics[scale=0.5]{figures/orcid.jpg}}}$, 
D. Piazza\inst{38}, 
G. Piotto\inst{21,39} $^{\href{https://orcid.org/0000-0002-9937-6387}{\includegraphics[scale=0.5]{figures/orcid.jpg}}}$, 
D. Pollacco\inst{12}, 
D. Queloz\inst{40,13} $^{\href{https://orcid.org/0000-0002-3012-0316}{\includegraphics[scale=0.5]{figures/orcid.jpg}}}$, 
R. Ragazzoni\inst{21,39} $^{\href{https://orcid.org/0000-0002-7697-5555}{\includegraphics[scale=0.5]{figures/orcid.jpg}}}$, 
N. Rando\inst{17}, 
H. Rauer\inst{14,41,42} $^{\href{https://orcid.org/0000-0002-6510-1828}{\includegraphics[scale=0.5]{figures/orcid.jpg}}}$, 
I. Ribas\inst{43,44} $^{\href{https://orcid.org/0000-0002-6689-0312}{\includegraphics[scale=0.5]{figures/orcid.jpg}}}$, 
N. C. Santos\inst{2,3} $^{\href{https://orcid.org/0000-0003-4422-2919}{\includegraphics[scale=0.5]{figures/orcid.jpg}}}$, 
D. Ségransan\inst{1} $^{\href{https://orcid.org/0000-0003-2355-8034}{\includegraphics[scale=0.5]{figures/orcid.jpg}}}$, 
A. E. Simon\inst{6,7} $^{\href{https://orcid.org/0000-0001-9773-2600}{\includegraphics[scale=0.5]{figures/orcid.jpg}}}$, 
M. Stalport\inst{27,26}, 
Gy. M. Szabó\inst{45,46} $^{\href{https://orcid.org/0000-0002-0606-7930}{\includegraphics[scale=0.5]{figures/orcid.jpg}}}$, 
N. Thomas\inst{6}, 
S. Udry\inst{1} $^{\href{https://orcid.org/0000-0001-7576-6236}{\includegraphics[scale=0.5]{figures/orcid.jpg}}}$, 
V. Van Grootel\inst{27} $^{\href{https://orcid.org/0000-0003-2144-4316}{\includegraphics[scale=0.5]{figures/orcid.jpg}}}$, 
J. Venturini\inst{1} $^{\href{https://orcid.org/0000-0001-9527-2903}{\includegraphics[scale=0.5]{figures/orcid.jpg}}}$, 
E. Villaver\inst{15,16}, 
N. A. Walton\inst{47} $^{\href{https://orcid.org/0000-0003-3983-8778}{\includegraphics[scale=0.5]{figures/orcid.jpg}}}$
}

\authorrunning{Akinsanmi et al.}

\institute{\label{inst:1} Observatoire astronomique de l'Université de Genève, Chemin Pegasi 51, 1290 Versoix, Switzerland \and
\label{inst:2} Instituto de Astrofisica e Ciencias do Espaco, Universidade do Porto, CAUP, Rua das Estrelas, 4150-762 Porto, Portugal \and
\label{inst:3} Departamento de Fisica e Astronomia, Faculdade de Ciencias, Universidade do Porto, Rua do Campo Alegre, 4169-007 Porto, Portugal \and
\label{inst:4} Space Research Institute, Austrian Academy of Sciences, Schmiedlstrasse 6, A-8042 Graz, Austria \and
\label{inst:5} INAF, Osservatorio Astrofisico di Torino, Via Osservatorio, 20, I-10025 Pino Torinese To, Italy \and
\label{inst:6} Weltraumforschung und Planetologie, Physikalisches Institut, University of Bern, Gesellschaftsstrasse 6, 3012 Bern, Switzerland \and
\label{inst:7} Center for Space and Habitability, University of Bern, Gesellschaftsstrasse 6, 3012 Bern, Switzerland \and
\label{inst:8} Department of Astronomy, Stockholm University, AlbaNova University Center, 10691 Stockholm, Sweden \and
\label{inst:9} Centre for Exoplanet Science, SUPA School of Physics and Astronomy, University of St Andrews, North Haugh, St Andrews KY16 9SS, UK \and
\label{inst:10} IMCCE, UMR8028 CNRS, Observatoire de Paris, PSL Univ., Sorbonne Univ., 77 av. Denfert-Rochereau, 75014 Paris, France \and
\label{inst:11} INAF, Osservatorio Astrofisico di Catania, Via S. Sofia 78, 95123 Catania, Italy \and
\label{inst:12} Department of Physics, University of Warwick, Gibbet Hill Road, Coventry CV4 7AL, United Kingdom \and
\label{inst:13} Cavendish Laboratory, JJ Thomson Avenue, Cambridge CB3 0HE, UK \and
\label{inst:14} Institute of Planetary Research, German Aerospace Center (DLR), Rutherfordstrasse 2, 12489 Berlin, Germany \and
\label{inst:15} Instituto de Astrofisica de Canarias, Via Lactea s/n, 38200 La Laguna, Tenerife, Spain \and
\label{inst:16} Departamento de Astrofisica, Universidad de La Laguna, Astrofísico Francisco Sanchez s/n, 38206 La Laguna, Tenerife, Spain \and
\label{inst:17} European Space Agency (ESA), ESTEC, Keplerlaan 1, 2201 AZ Noordwijk, The Netherlands \and
\label{inst:18} Admatis, 5. Kandó Kálmán Street, 3534 Miskolc, Hungary \and
\label{inst:19} Depto. de Astrofisica, Centro de Astrobiologia (CSIC-INTA), ESAC campus, 28692 Villanueva de la Cañada (Madrid), Spain \and
\label{inst:20} Université Grenoble Alpes, CNRS, IPAG, 38000 Grenoble, France \and
\label{inst:21} INAF, Osservatorio Astronomico di Padova, Vicolo dell'Osservatorio 5, 35122 Padova, Italy \and
\label{inst:22} Institute of Optical Sensor Systems, German Aerospace Center (DLR), Rutherfordstrasse 2, 12489 Berlin, Germany \and
\label{inst:23} Université de Paris Cité, Institut de physique du globe de Paris, CNRS, 1 Rue Jussieu, F-75005 Paris, France \and
\label{inst:24} Centre for Mathematical Sciences, Lund University, Box 118, 221 00 Lund, Sweden \and
\label{inst:25} Aix Marseille Univ, CNRS, CNES, LAM, 38 rue Frédéric Joliot-Curie, 13388 Marseille, France \and
\label{inst:26} Astrobiology Research Unit, Université de Liège, Allée du 6 Août 19C, B-4000 Liège, Belgium \and
\label{inst:27} Space sciences, Technologies and Astrophysics Research (STAR) Institute, Université de Liège, Allée du 6 Août 19C, 4000 Liège, Belgium \and
\label{inst:28} Centre Vie dans l’Univers, Faculté des sciences, Université de Genève, Quai Ernest-Ansermet 30, 1211 Genève 4, Switzerland \and
\label{inst:29} Leiden Observatory, University of Leiden, PO Box 9513, 2300 RA Leiden, The Netherlands \and
\label{inst:30} Department of Space, Earth and Environment, Chalmers University of Technology, Onsala Space Observatory, 439 92 Onsala, Sweden \and
\label{inst:31} Dipartimento di Fisica, Università degli Studi di Torino, via Pietro Giuria 1, I-10125, Torino, Italy \and
\label{inst:32} Department of Astrophysics, University of Vienna, Türkenschanzstrasse 17, 1180 Vienna, Austria \and
\label{inst:33} Institute for Theoretical Physics and Computational Physics, Graz University of Technology, Petersgasse 16, 8010 Graz, Austria \and
\label{inst:34} Konkoly Observatory, Research Centre for Astronomy and Earth Sciences, 1121 Budapest, Konkoly Thege Miklós út 15-17, Hungary \and
\label{inst:35} ELTE E\"otv\"os Lor\'and University, Institute of Physics, P\'azm\'any P\'eter s\'et\'any 1/A, 1117 Budapest, Hungary \and
\label{inst:36} Institut d'astrophysique de Paris, UMR7095 CNRS, Université Pierre \& Marie Curie, 98bis blvd. Arago, 75014 Paris, France \and
\label{inst:37} Astrophysics Group, Lennard Jones Building, Keele University, Staffordshire, ST5 5BG, United Kingdom \and
\label{inst:38} Physikalisches Institut, University of Bern, Gesellschaftsstrasse 6, 3012 Bern, Switzerland \and
\label{inst:39} Dipartimento di Fisica e Astronomia "Galileo Galilei", Università degli Studi di Padova, Vicolo dell'Osservatorio 3, 35122 Padova, Italy \and
\label{inst:40} ETH Zurich, Department of Physics, Wolfgang-Pauli-Strasse 2, CH-8093 Zurich, Switzerland \and
\label{inst:41} Zentrum für Astronomie und Astrophysik, Technische Universität Berlin, Hardenbergstr. 36, D-10623 Berlin, Germany \and
\label{inst:42} Institut fuer Geologische Wissenschaften, Freie Universitaet Berlin, Maltheserstrasse 74-100,12249 Berlin, Germany \and
\label{inst:43} Institut de Ciencies de l'Espai (ICE, CSIC), Campus UAB, Can Magrans s/n, 08193 Bellaterra, Spain \and
\label{inst:44} Institut d’Estudis Espacials de Catalunya (IEEC), Gran Capità 2-4, 08034 Barcelona, Spain \and
\label{inst:45} ELTE E\"otv\"os Lor\'and University, Gothard Astrophysical Observatory, 9700 Szombathely, Szent Imre h. u. 112, Hungary \and
\label{inst:46} HUN-REN–ELTE Exoplanet Research Group, Szent Imre h. u. 112., Szombathely, H-9700, Hungary \and
\label{inst:47} Institute of Astronomy, University of Cambridge, Madingley Road, Cambridge, CB3 0HA, United Kingdom \and
\label{inst:48} CFisUC, Departamento de Física, Universidade de Coimbra, 3004-516 Coimbra, Portugal
}

   \date{Received 06 October 2023; accepted 05 February 2024 }

  \abstract
   {Ultra-hot Jupiters present a unique opportunity to understand the physics and chemistry of planets, their atmospheres, and interiors at extreme conditions. WASP-12\,b stands out as an archetype of this class of exoplanets, with a close-in orbit around its star that results in intense stellar irradiation and tidal effects.}
   {The goals are to measure the planet's tidal deformation, atmospheric properties, and also to refine its orbital decay rate.}
   {We performed comprehensive analyses of the transits, occultations, and phase curves of \plname by combining new \cheops observations with previous \tess and \spitzer data. The planet was modeled as a triaxial ellipsoid parameterized by the second-order fluid Love number of the planet, $h_2$, which quantifies its radial deformation and provides insight into the interior structure.}
  {We measured the tidal deformation of \plname and estimated a Love number of \hf=$1.55_{-0.49}^{+0.45}$ (at 3.2$\sigma$) from its phase curve. We measured occultation depths of $333\pm24$\,ppm and $493\pm29$\,ppm in the \cheops and \tess bands, respectively, while the nightside fluxes are consistent with zero, and also marginal eastward phase offsets. Our modeling of the dayside emission spectrum indicates that \cheops and \tess probe similar pressure levels in the atmosphere at a temperature of $\sim$2900\,K. We also estimated low geometric albedos of $A_g=0.086\pm0.017$  and $A_g=0.01\pm0.023$ in the \cheops and \tess passbands, respectively, suggesting the absence of reflective clouds in the high-temperature dayside of the planet. The \cheops occultations do not show strong evidence for variability in the dayside atmosphere of the planet at the median occultation depth precision of 120\,ppm attained. Finally, combining the new \cheops timings with previous measurements refines the precision of the orbital decay rate by 12\% to a value of –30.23$\pm$0.82 ms/yr, resulting in a modified stellar tidal quality factor of $Q'_{\star}=1.70\pm0.14\times 10^5$.}
   {\plname becomes the second exoplanet, after WASP-103b, for which the Love number has been measured from the effect of tidal deformation in the light curve. However, constraining the core mass fraction of the planet requires measuring $h_2$ with a higher precision. This can be achieved with high signal-to-noise observations with JWST since the phase curve amplitude, and consequently the induced tidal deformation effect, is higher in the infrared.
   }

   \keywords{Planetary systems – Planets and satellites: individual: WASP-12\,b, atmospheres, interiors – Techniques: photometric}

   \maketitle
%
\section{Introduction}

Ultra-hot Jupiters (UHJs) orbit very close to their host stars and are subjected to immense tidal forces and irradiation which impact the orbital, atmospheric, and geometric characteristics of the planets. Depending on the stellar properties, the strong tidal interaction may cause the orbits of the planets to become circular and coplanar (zero eccentricity and obliquity), while their rotation rates and orbital periods may become synchronized \citep[tidal locking;][]{Hut1980}. These effects can impact the atmospheric circulation and also result in tidal deformation of the planet's shape in response to the perturbing force \citep{corr13,Correia2014}. Another consequence of the tidal interaction is the shrinkage of the planetary orbit (tidal decay) due to loss of angular momentum to the star if the planetary orbital period is not synchronized with the stellar rotation period. 

The intense irradiation received by UHJs results in extremely high dayside atmospheric temperatures such that molecular water (H$_2$O) is thermally unstable in favor of atomic hydrogen \citep{2018A&A...614A...1W}. The hot daysides of UHJs generally favor the atomic form rather than the molecular form (e.g., Si/SiO or Mg/MgH). Other metal elements may appear in their ionized form, such as Na$^+$, K$^+$, but also in the less abundant Ti$^+$ and Al$^+$ \citep{2019A&A...626A.133H,2021A&A...648A..80H}. Therefore, UHJs are unique laboratories to study the physics and chemistry of planets at extreme conditions, shedding valuable insights into their orbital evolution, atmospheres, and interiors.

WASP-12\,b stands out as one of the most irradiated UHJs, with an orbital period of only 1.09\,days around a G0 star of ${T_\mathrm{eff}}$=6300\,K \citep{hebb-2009-wasp-12}. Separated from the star by less than 3 stellar radii, the planet is exposed to strong tidal forces and irradiation, which makes it an attractive target for characterization via transit, eclipse, and phase-curve observations. WASP-12\,b is indeed one of the most extensively studied exoplanets, with numerous observations ranging from ultraviolet to infrared wavelengths, revealing remarkable properties of the planet. The planet is inflated with a large radius of $1.9R_{\mathrm{Jup}}$  likely due to high stellar irradiation. Evidence suggests that it is undergoing atmospheric mass loss with gas outflowing toward the star \citep{Fossati2010MetalsWASP-12b}. The proximity of its orbit to the Roche limit of its star makes it one of the few exoplanets where the tidal distortion of the planet can be probed from its light curve \citep{correia14, akin19}. It is also the only exoplanet that has been observationally confirmed to be spiraling into the star due to tidal orbital decay  \citep{Yee2019TheDecaying}. Secondary eclipse observations of WASP-12\,b have resulted in discrepant eclipse depth measurements at various passbands which could be indicative of variability in the dayside atmosphere \citep{Hooton2019StormsWASP-12b}. Previous work on \plname found evidence for water absorption in its terminator \citep{stevenson2014,Kreidberg2015}, whereas the dayside spectrum showed no signs of water \citep{swain-2013-wasp-12}, supporting previous gas-phase modeling results \citep[e.g.,][]{2019A&A...626A.133H}.

In this paper, we report on the transit, occultation, and phase curve observations of \plname by the CHaracterizing ExOplanet Satellite \citep[\cheops;][]{Benz2021}. Previous \cheops observations \citep[e.g.,][]{Lendl2020, Barros2022, deline2022, Hooton2021, Ehrenreich2023ACHEOPS} have shown its remarkable photometric capability in characterizing exoplanets. Here, we analyze the \cheops observations alongside archival data of \stname to characterize the shape, orbit, and atmosphere of the planet. In Section\,\ref{sect:observations}, we described the observations obtained by the different instruments, and also the pre-processing of the datasets. We also derived the stellar properties and summarized the theoretical background on planetary tidal deformation. The analyses of the datasets are detailed in Section\,\ref{sect:analysis}. In Section\,\ref{sect:tidal_deformation}, we use the ellipsoidal planet model to probe the tidal deformation of the planet in the joint phase curve and transit observations. We characterize the atmosphere of the planet in Section\,\ref{sect:atmosphere}: modeling its emission spectrum and probing for variability in its dayside atmosphere. In Section\,\ref{sect:tidal_decay}, we use our \cheops transit timing measurements along with published ones to refine the tidal decay timescale of the planet and the stellar tidal dissipation. Finally, we summarize the main results of our work in Section\,\ref{sect:conclusion}.

\section{Observations and system properties}
\label{sect:obs_sys}
\subsection{Observations}
\label{sect:observations}
\stname has been observed by several space- and ground-based instruments, capturing the transit, eclipse, and also the phase curve of the planet,  \plname. In this paper, we analyze new observations from \cheops in addition to existing space-based observations from \tess and \spitzer. Although \texttt{HST/WFC3} observations of \plname are also available, we excluded them due to poor transit ingress and/or egress coverage that makes it challenging to probe the planetary deformation.

\subsubsection{CHEOPS}
We obtained 47 visits of \stname spanning 3 observation seasons between 2022–11–02 and 2022–12–24 using \cheops as part of the Guaranteed Time Observations (See observation log in Table\,\ref{tab:observations}).  The visits, identified by unique file keys, consist of 21 transits, 25 occultations, and half of a phase curve of \plname, all taken at an exposure time of 60\,s. The phase curve visit lasted for 24\,hrs, starting before an occultation and ending after transit. The visit durations of the transits and occultations range between 7.1 – 12.6\,hrs, capturing significant baselines before and after the transits/occultations (see Figs.\,\ref{fig:transit_obs} and \ref{fig:occulation_obs}). Due to the short orbital period of the planet, the transit and occultation visits also fortuitously combine to construct a phase curve for the planet. In some cases, consecutive observations of transits and occultations result in half or full phase curves. 

Due to the low-Earth orbit of \cheops, its line of sight is often interrupted by Earth occultations of the target or spacecraft passages through the South Atlantic Anomaly resulting in data gaps. For our observations of \stname, this resulted in light curve efficiencies between 49 and 63\%. \texttt{CHEOPS} data are automatically processed by the official Data Reduction Pipeline \citep[DRP version 13;][]{Hoyer2020} which performs aperture photometry after calibration of the images and correcting for instrumental and environmental effects. The DRP provides light curves extracted with different aperture radii. 
Point-spread function (PSF) photometry can also be extracted using the \texttt{PIPE}\footnote{\href{https://github.com/alphapsa/PIPE}{https://github.com/alphapsa/PIPE}} package developed specifically for \cheops data.

In our analyses of \stname, we used the PSF-extracted light curves which are less sensitive to contamination from background stars \citep[see e.g.,][]{Morris2020, brandeker2022, Delrez2023RefiningTESS}. The resulting light curves have a lower scatter than the DRP apertures in all visits. Data points that were flagged to have poor photometry (e.g., due to cosmic ray hits or bad pixels) were discarded. We further removed points with high background ($BG>3$ times the median background level) where the correlation with the flux becomes nonlinear due to scattered light from the moon or the Earth's limb. Finally, a 15-point moving median filter was used to eliminate points $>$5 times the median absolute deviation (MAD). In total, 987 points were discarded corresponding to 6.2\% of the data points across all visits.  

The Nadir-locked orientation of \cheops, as it orbits around the Earth, causes its field of view to rotate around the target. Combined with the irregular shape of the \cheops PSF, this leads to time-variable flux contamination in the aperture that is correlated with the spacecraft's roll-angle. Thus, it is usually necessary to decorrelate against the roll-angle when analyzing \cheops data \citep[e.g.,][]{Lendl2020, Morris2021a, Barros2022}. Spacecraft pointing jitter can also result in flux trends that can be accounted for by decorrelating against the X and Y centroid positions of the target PSF on the CCD. Furthermore, \cheops observations can feature ramp effects at the beginning of each visit caused by the thermal settling of the telescope as it adjusts to a new target position. The ramp is accounted for by decorrelating the flux against the deviation of telescope tube temperature from the median value $\Delta T_{\mathrm{tube}}$ \citep[see e.g.,][]{Morris2021a, deline2022}. Data points ($\sim$45) in the first orbit of visits 44, 45, and 47 were removed as they featured a strong, nonlinear increase in the telescope temperature.

We model the systematic trends in each \cheops visit using spline decorrelations\footnote{implemented with the \texttt{scipy.interpolate.LSQUnivariateSpline} and \texttt{scipy.interpolate.LSQBivariateSpline} python classes.} against the roll-angle, background flux, telescope tube temperature, and the $X$ and $Y$ centroid positions. The spline fit is performed simultaneously with the fit of the astrophysical model (\S\ref{sect:LC_model}) and involves successively fitting splines to the residuals of the astrophysical model and then evaluating the likelihood of the joint model. First, we performed a 2D spline fit of the residual against $BG$ and $\Delta T_{\mathrm{tube}}$. The resulting residual is then used for another 2D spline fit against the $X$ and $Y$ centroid positions. Since the flux trends with these variables are approximately linear, the 2D spline functions are defined with a single degree and knot in each dimension. Finally, we model the roll-angle trend with a 1D cubic spline fit with knots every 18$\degree$.

\subsubsection{TESS}
\tess observed \stname with 2-minute cadence in sectors 20, 43, 44, and 45 with a span of almost 2\,years between December 2019 and December 2021. Across the four sectors, \tess observed 74 transits and occultations. Details of the \tess dataset are given in Table\,\ref{tab:tess_obs}. We utilized the Pre-search Data Conditioning Single Aperture Photometry (PDCSAP) light curve data produced by the Science Processing Operations Center (SPOC) pipeline which has been corrected for known instrumental systematics and contamination \citep{stumpe2012,smith2012}. The light curves from these sectors were recently published by \cite{Wong2022-WASP-12} where the transits, occultations, phase curve, and transit timings for WASP-12b were analyzed. We also analyzed these light curves to complement our \cheops observations. 

The lightcurves were downloaded from the Mikulski Archive for Space Telescope (MAST) archive using the \texttt{lightkurve}\textit{ python} package \citep{lightkurve2018}. Data points flagged by the SPOC pipeline were removed after which a 15-point moving median filter was used to remove points $>$5$\times$MAD. We separated the light curves into segments by the time of the momentum dumps and removed any strong flux ramps at the beginning of data segments. In fitting the astrophysical model, we simultaneously account for long-term temporal trends in each data segment using a cubic spline with knots every 3\,days so as to preserve the phase variation within an orbital period.

\subsubsection{Spitzer}
We also analyzed archival \spitzer data in the 3.6$\mu$m and 4.5$\mu$m channels of the InfraRed Array Camera (IRAC) that have already been published \citep{cowan2012-wasp-12, bell-2019-WASP-12}. These observations consist of 2 phase curves in each channel acquired in 2010 (PID 70060, PI P. Machalek) and 2013 (PID 90186, PI K. Todorov). We downloaded the data from the Spitzer Heritage Archive\footnote{\url{http://sha.ipac.caltech.edu}}.

The reduction and analysis of these datasets are similar to \citet{Demory:2016b}, where we modeled the IRAC intra-pixel sensitivity \citep{Ingalls:2012} using a modified implementation of the BLISS (BiLinearly-Interpolated Sub-pixel Sensitivity) mapping algorithm \citep{stevenson2012_Bliss_mapping}. In addition to the BLISS mapping (BM), our baseline model includes a linear function of the Point Response
Function’s (PRF) FWHM along the $x$ and $y$ axes, which significantly reduces the level of correlated noise as shown in previous studies \citep[e.g.,][]{Demory2016, Demory:2016b, Mendonca:2018, Barros2022,jones2022}. This baseline model (BM + PRF FWHM) does not include time-dependent parameters. We implemented this instrumental model in a Markov Chain Monte Carlo (MCMC) framework already presented in the literature (\citealt{Gillon:2012a}). We included all data described in the paragraph above in the same fit. We ran two chains of 200,000 steps each to determine baseline corrected light curves at 3.6 and 4.5 $\mu$m that were used subsequently.

Previous analyses of these datasets by \citet{cowan2012-wasp-12} and \citet{bell-2019-WASP-12} reported an anomalous phase modulation (with a periodicity of half the planet's orbital period) in the 4.5$\mu$m data but not at 3.6$\mu$m. If the anomalous modulation is due to the tidal deformation of the planet, \citet{cowan2012-wasp-12} estimated that the substellar axis would have to be at least 1.5 times longer than the polar axis. However, the effect of such a large deformation is not supported by the transit. \citet{bell-2019-WASP-12} instead proposed that the anomalous signal at 4.5$\mu$m may be due to heated CO emission from gas outflowing from the planet. \citet{bell-2019-WASP-12} also found that the measured hotspot offset of the 3.6$\mu$m phase curves significantly changed from eastward in 2010 to westward in 2013. As the unexpected phase curve features in these datasets make them difficult to combine, we chose to use only the transit regions of the \spitzer data in our analysis (0.2\,d before and after mid-transit). Similar to the \cheops and \tess datasets, we remove outlier points $>$5$\times$MAD with a 15-point moving median filter. Details of the \spitzer datasets are given in Table\,\ref{tab:spitzer_obs}.

\subsection{The host star}
\subsubsection{Stellar parameters}
\label{sect:star_pars}
\begin{table}[t]
\centering
\caption{Properties of the star WASP-12 system}
\label{tab:stellar_table}
\resizebox{0.5\textwidth}{!}{%
\begin{tabular}{llll}
\hline\hline
Parameter name        & Symbol                       & Value       & Source    \\ \hline
Effective temperature & $T_{\mathrm{eff}}$\,{[}K{]}  & $6313\pm52$              &  This work\\
Surface gravity       & $\log{g}$ {[}dex{]}          & $4.37\pm0.12$             & This work \\
Metallicity           & {[}Fe/H{]} {[}dex{]}         & $0.21\pm0.04$            & This work\\
Stellar radius        & $R_{\star}$ {[}$R_{\odot}${]} & $1.734\pm0.022$          &  This work\\
Stellar mass          & $M_{\star}$ {[}$M_{\odot}${]} & $1.422_{-0.069}^{+0.077}$ &  This work\\
Stellar age           & $t_{\star}$\,[Gyr]                         & $2.3\pm0.5$              &  This work\\
Planet Mass           & $M_{p}$ {[}$M_{J}${]}        & $1.470\pm0.073$          &  \citet{Collins2017TRANSITPLANETS}\\
RV semi-amplitude     & $K_{\mathrm{RV}}$ [m/s]       & $226.4\pm4.1$          &  \citet{Collins2017TRANSITPLANETS}\\\hline
\end{tabular}%
}
\end{table}

To facilitate our analysis of the observations, we refined the stellar parameters of \stname (V\,=\,11.5) as shown in Table\,\ref{tab:stellar_table}. The spectroscopic stellar parameters ($T_{\mathrm{eff}}$, $\log g$, microturbulence, [Fe/H]) were originally taken from a previous version of SWEET-Cat \citep{Santos-13, Sousa-18}. The spectroscopic parameters for WASP-12 were estimated with the ARES+MOOG methodology where we used the latest version of ARES\footnote{The latest version, ARES v2, can be downloaded at https://github.com/sousasag/ARES} \citep{Sousa-07, Sousa-15} to consistently measure the equivalent widths (EW) of selected iron lines on the spectrum of WASP-12. The list of iron lines is the same as the one presented in \citet[][]{Sousa-08}. In this analysis, we used a minimization process to find the ionization and excitation equilibrium to converge on the best set of spectroscopic parameters. This process makes use of the ATLAS grid of stellar model atmospheres \citep{Kurucz-93} and the radiative transfer code MOOG \citep{Sneden-73}. More recently the same methodology was applied on a combined HARPS–N spectrum where we derived completely consistent spectroscopic stellar parameters ($T_{\mathrm{eff}}\,=\,6301\pm 64$\,K, $\log g\,=\,4.26\pm0.10$ dex, and [Fe/H] = $0.18\pm0.04$ dex; \citealt[][]{Sousa-21}). We also derived a more accurate trigonometric surface gravity using recent GAIA data following the same procedure as described in \citet[][]{Sousa-21} which provided a consistent value when compared with the spectroscopic surface gravity (4.23 $\pm$ 0.01 dex).

We determined the radius of WASP-12 using the infrared flux method (IRFM) with an MCMC approach \citep{Blackwell1977,Schanche2020}. We conducted a comparison between observed and synthetic broadband photometry to determine the stellar effective temperature and angular diameter that is converted to the stellar radius with knowledge of the target's parallax. We constructed the spectral energy distributions (SEDs) of WASP-12 using the results of our spectral analysis above and the ATLAS stellar atmosphere models. We then produce synthetic photometry in the {\it Gaia}, 2MASS, and {\it WISE} passbands that are compared to {\it Gaia} G, G$_{\rm BP}$, and G$_{\rm RP}$, 2MASS J, H, and K, and {\it WISE}\ W1 and W2 broadband fluxes from the most recent data releases \citep{Skrutskie2006,Wright2010,GaiaCollaboration2022}. Using the offset-corrected {\it Gaia} DR3 parallax, we obtained the stellar radius that is reported in Table\,\ref{tab:stellar_table}.

We used $T_{\mathrm{eff}}$, [Fe/H], and $R_{\star}$ along with their uncertainties to determine the stellar mass $M_{\star}$ and age $t_{\star}$ by employing two different stellar evolution models. In fact, a first pair of mass and age values ($M_{\star,1}$, $t_{\star,1}$) were computed by the isochrone placement algorithm \citep{bonfanti15,bonfanti16} that interpolates the input values within precomputed grids of PARSEC\footnote{\textsl{PA}dova and T\textsl{R}ieste \textsl{S}tellar \textsl{E}volutionary \textsl{C}ode: \url{http://stev.oapd.inaf.it/cgi-bin/cmd}} v1.2S \citep{marigo17} isochrones and tracks. A second pair ($M_{\star,2}$, $t_{\star,2}$), instead, was computed via the CLES \citep[Code Liègeois d'Évolution Stellaire][]{scuflaire08} code, which generates the best-fit evolutionary track according to the provided input and the Levenberg-Marquadt minimization scheme \citep{salmon21}.
We finally merged the two pairs of outcomes after successfully checking their mutual consistency through the $\chi^2$-based criterion as described in \citet{bonfanti21} and we obtained $M_{\star}=1.422_{-0.069}^{+0.077}\,M_{\odot}$ and $t_{\star}=2.3\pm0.5$ Gyr, consistent with literature values.

\subsubsection{Stellar limb darkening}
\label{sect:limb_darkening}
Accurate modeling of stellar limb darkening (LD) is important when analyzing exoplanetary transits in order to obtain unbiased transit parameters, and also to measure higher-order effects such as tidal deformation \citep{Espinoza2015, akin19}. 
The theoretical LD profile of different stars can be obtained from stellar atmosphere models that compute the stellar intensities as a function of the foreshortening angle $\mu$ measured from the limb to the center of the stellar disk. The most widely used stellar libraries for this purpose are the \textsc{PHOENIX} \citep{husser} and ATLAS \citep{Kurucz-93} stellar models. Since they are both theoretical models, they may not always provide an accurate representation of the actual stellar intensity profile \citep[see e.g.,][]{Espinoza2015, Patel2022EmpiricalTESS}. Indeed, both libraries predict slightly different intensity profiles for the same star in the same passband, making it difficult to select one over the other. However, since these libraries represent our current knowledge of stellar atmospheres, they may still be used to put useful priors on the stellar limb darkening profiles. Obtaining priors from these models can also be beneficial to multi-passband observations since the derived priors will ensure that the LD profile in all passbands relates to the same star.

First, we compute stellar intensity profiles for each passband using the \texttt{LDCU} python package\footnote{\href{https://github.com/delinea/LDCU}{https://github.com/delinea/LDCU}} which queries both stellar libraries using the stellar parameters ($T_{\mathrm{eff}}$, $\log g$, and [Fe/H]) given in Table\,\ref{tab:stellar_table}. Following \citet{claret}, the obtained intensity profile computed from each library, and in each passband, is represented by 100 interpolated points (evenly spaced in $\mu$). The uncertainties in the stellar parameters are accounted for by generating 10000 intensity profile samples using random stellar parameter sets drawn from their normal distribution. The median and standard deviation of the profiles are then computed. Thus, for each passband, we have a median intensity profile from each stellar library with error bars at each $\mu$ point. 

Different parametric limb darkening laws can be adopted to approximate the model intensity profile to derive limb darkening coefficients (LDCs) that can be used in transit analyses. In this work, we adopted the power-2 LD law \citep{Hestroffer1997} parameterized by the LDCs, $c$ and $\alpha$. The power-2 law has been shown to be superior to other two-parameter laws in modeling intensity profiles generated by stellar atmosphere models \citep{morello, Claret_southworth2022_power2_ldcs}. The power-2 LD law is only surpassed by the four-parameter law which is difficult to use in transit model fitting due to the higher number of parameters and the strong correlations between them, which can lead to nonphysical intensity profiles.

Similar to \citet{Barros2022}, we leveraged the computed model intensity profiles from the two libraries to derive priors on the LDCs. 
We obtain LDCs in each passband by fitting the preferred power-2 law to the corresponding combined \textsc{PHOENIX} and \textsc{ATLAS} model intensity profiles. The $1\sigma$ uncertainties of the obtained LDCs are inflated such that the allowed parameter space encompasses both intensity profiles and associated $1\sigma$ uncertainties. This approach is illustrated in Fig.\,\ref{fig:ldc_plot}a and the derived LDC priors for the passbands are reported in Table\,\ref{tab:ldc_priors}. Using the derived LDCs as priors allows the transit fit to determine the best-fit limb darkening profile without being too restricted to the predictions of either library.  Alternative approaches to modeling limb darkening using the intensity profiles are presented in Section\,\ref{sect:ldc_impact}.

\subsection{The planet: Tidal deformation}
\label{sect:planet_deformation}
WASP-12\,b orbits so close to its host star that it is predicted to be one of the most tidally deformed planets \citep{akin19,Hellard2019,Berardo2022TidalCompositions}. The deformation of a planet in response to perturbing forces depends on its interior structure and can be quantified by the second-degree Love number for radial displacement \hf  \citep{love, Kellermann2018}. For a fluid planet, \hf is related to the tidal Love number $k_2$ (as \hf = 1 + $k_2$) which measures the distribution of mass within the planet. Therefore, measurement of \hf from the detection of tidal deformation allows constraining the planet's interior structure \citep{kramm11,kramm12}. The relationship between the tidal response of a planet and its core mass has been investigated in previous studies \citep[e.g.,][]{Batygin2009,Ragozzine2009,kramm11} showing that an incompressible fluid planet with a homogeneous interior mass distribution will have the highest \hf value of 2.5. However, the value of \hf decreases as a planet becomes more centrally condensed (more mass at the core), since the presence of a massive core reduces the response of a planet to the perturbing potential \citep{Leconte2011}. For instance, the lower measured Love number of 1.39 for Saturn \citep{LAINEY2017} reflects a core mass fraction higher than that of Jupiter which has a Love number of 1.565 \citep{Durante2020}.  

The shape of a tidally deformed planet can be described by a triaxial ellipsoid (with axes $r_1,\,r_2,\,r_3$) where $r_1$ is the planet radius oriented along the star-planet (substellar) axis, $r_2$ is along the orbital direction (dawn-dusk axis), and $r_3$ is along the polar axis. The volumetric radius of the ellipsoid is given by \Rv{}$~=~(r_1\,r_2\,r_3)^{1/3}$. According to the analytical shape model formulated in \citet{correia14}, the axes of the ellipsoid follow the relation\footnote{This relation also satisfies the condition in the linear theory of figures \citep{Zharkov1978PhysicsInteriors,Dermott1979TidalPlanets} that a synchronously rotating body in hydrostatic equilibrium subjected to both rotation and tidal forces deforms such that $(r_2 - r_3)/(r_1 - r_3) = 1/4$}: 
\begin{align}
    \label{eqn:ell_axis}
    r_2 = \Rv(1-2q/3),~~ r_1\,=\,r_2(1+3q),~~\text{and}~~ r_3\,=\,r_2(1-q)
\end{align}

\noindent where $q$ is an asymmetry parameter defined as
\begin{equation}
    q  = \frac{\hf}{2Q_{\mathrm{M}}}\left(\frac{\Rv{}}{a}\right)^3\,.
    \label{eqn:asym_par}
\end{equation}

\noindent Therefore the axes $r_1,\,r_2,\,r_3$ of the deformed planet depend on the Love number \hf, the proximity of the planetary orbit to the star $a$, the planet-to-star mass ratio $Q_{\mathrm{M}}=M_p/M_{\star}$, and the volumetric radius of the planet \Rv. 

The nonspherical shape of the ellipsoidal planet causes the projected cross-sectional area to vary as it rotates with orbital phase. The projected area as a function of orbital phase angle ($\phi$\,=\,2$\pi$\,phase) is given \citep[e.g., by][]{Leconte2011a} as 
\begin{equation}
    A(\phi) = \pi\sqrt{r_1^2r_2^2\cos^2i + r_3^2\sin^2i\,\left(r_1^2\sin^2{\phi} + r_2^2\cos^2{\phi}\right)},
\label{eqn:varying_area}
\end{equation}

\noindent where $i$ is the orbital inclination of the planet. The projected area can be calculated using Eqs. (\ref{eqn:ell_axis}) and (\ref{eqn:asym_par}), given the values of \hf, \Rv, and $Q_M$. The ellipsoid projects its maximum area at quadrature, while the minimum area is projected at mid-transit (and mid-occultation). The \textit{effective} planetary radius at mid-transit ($\phi=0$) can be obtained from Eq.\,(\ref{eqn:varying_area}) as $R_p^{\mathrm{eff}}$\,=\,$\sqrt{A_{\phi=0}/\pi}$. The varying planet projection can lead to anomalies in the transit light curve \citep{correia14,akin19} and also in the shape of the full-orbit phase curve \citep{Leconte2011, Akinsanmi2023OnCurves} when compared to a spherical planet. Measuring the deviation of the planet's shape from sphericity in high-precision light curves can thus provide a measurement of the Love number.

\section{Light curve analysis}
\label{sect:analysis}
In this section, we describe the analytical light curve model that we use to model the phase curves, transits, and occultations of \plname. We further describe our fitting methodology for each analysis.

\subsection{Light curve model}
\label{sect:LC_model}
We adopt an analytical light curve model composed of the transit ($F_{\mathrm{tra}}$) and occultation ($F_{\mathrm{occ}}$) signals, the phase variation signal by the planet ($F_p$: due to reflection and thermal emission from the atmosphere), and also the phase variation signal by the star ($F_{\star}$: due to ellipsoidal distortion of the star by the planet and Doppler beaming of the stellar light). The total phase curve model is given as a function of the orbital phase angle as:
\begin{equation}
    F(\phi) = F_{\mathrm{tra}}\, F_{\star}(\phi)~+~F_{\mathrm{occ}}\, F_p(\phi).
    \label{eqn:pc_flux}
\end{equation}

\noindent Given the expected deformed shape of \plname, an adequate light curve model should account for the deformation in the relevant component signals \citep{Akinsanmi2023OnCurves}. We describe the components of Eq.\,(\ref{eqn:pc_flux}) in the following sections.

\subsubsection{Stellar phase variation model}
\label{sect:stellar_vary}
The flux from the star $F_{\star}$ varies as a function of phase as
\begin{equation}
\begin{aligned}
    F_{\star}(\phi) &= 1 + F_\mathrm{EV} + F_\mathrm{DB}\\
                    &= 1 + A_{\mathrm{EV}}(1 - \cos{2\phi}) + A_{\mathrm{DB}}\sin{\phi}
    \label{eqn:st_flux},
\end{aligned}
\end{equation}

\noindent such that $F_{\star}$ is unity at mid-transit and mid-eclipse. The value of $F_{\star}$ at other phases depends on the ellipsoidal variation $F_{\mathrm{EV}}$ and Doppler beaming $F_{\mathrm{DB}}$ signals which have semi-amplitudes,  $A_{\mathrm{EV}}$ and $A_{\mathrm{DB}}$  respectively given \citep[e.g., in][]{Loeb2003PeriodicCompanions,Esteves2013OPTICALEXOPLANETS,Shporer2017-PCreview} by:
\begin{align}
    A_{\mathrm{EV}} &= \alpha_{\mathrm{EV}}\,Q_{\mathrm{M}}\left(\frac{a}{R_{\star}}\right)^{-3}\sin^2{i},\label{eqn:A_EV}\\
    A_{\mathrm{DB}} &= \alpha_{\mathrm{DB}}\,\frac{K_{\mathrm{RV}}}{c}
    \label{eqn:A_DB}
\end{align}
\noindent where $Q_{\mathrm{M}}$ is a again the planet-to-star mass ratio, $a/R_{\star}$ is the semi-major axis scaled by the stellar radius, $i$ is the orbital inclination, $K_{\mathrm{RV}}$ is the radial velocity (RV) semi-amplitude, and $c$ is the speed of light. The coefficient $\alpha_{\mathrm{EV}}$ depends on the linear limb darkening coefficient $u$, and gravity darkening coefficient $g$ as
\begin{equation}
    \alpha_{\mathrm{EV}} = 0.15\frac{(15+u)(1+g)}{3-u},
    \label{eqn:alpha_EV}
\end{equation}
\noindent while the coefficient $\alpha_{\mathrm{DB}}$ depends on the stellar flux $\mathcal{F}_{\lambda}$ and passband transmission $\mathcal{T}_{\lambda}$ at wavelength $\lambda$ as
\begin{equation}
    \alpha_{\mathrm{DB}} = \frac{\int\left( 5+\frac{d\ln \mathcal{F}_{\lambda}}{d\ln \lambda} \right)\lambda \mathcal{F}_{\lambda}\mathcal{T}_{\lambda} d\lambda}{\int\lambda \mathcal{F}_{\lambda}\mathcal{T}_{\lambda} d\lambda}.
    \label{eqn:alpha_db}
\end{equation}

\subsubsection{Transit and occultation models}
\label{sect:trans_occ}
The transit, $F_{\mathrm{tra}}$, and occultation, $F_{\mathrm{occ}}$, signals are generated using the \texttt{ellc} transit tool \citep{maxted} which allows modeling the planet shape as a sphere or as an ellipsoid parameterized by the Love number as implemented in \citet{akin19}. The ellipsoidal planet model parameters are the same as the usual spherical planet model except that the spherical planet radius $R_p$ is replaced by the volumetric radius \Rv\footnote{\Rv of an ellipsoidal planet is always larger than the spherical planet transit radius since only a small part of the long axis is projected during transit. However, for the spherical planet, \Rv is equivalent to the spherical planet transit radius.}, and the addition of \hf and $Q_M$.

\subsubsection{Planetary phase variation model}
\label{sect:planet_vary}
Although the flux from the planet ($F_p$) is composed of both reflected light and thermal emission from the atmosphere, the degeneracy between both components makes it challenging to model them simultaneously \citep[see e.g.,][]{Lendl2020, deline2022, parvieinen2022}. Following recent optical phase curve studies \citep[e.g.,][]{shporer-2019-WASP18, wong2021_tess2ndyearPCs, Daylan2021TESSCurve}, we model the total planetary phase variation using a sinusoidal function given by:
\begin{equation}
    F_{p}(\phi)  =  \left(F_{\mathrm{max}} - F_{\mathrm{min}}\right) \, \frac{1 - \cos{(\phi + \delta)}}{2} + F_{\mathrm{min}},
    \label{eqn:planet_flux}
\end{equation}
where $F_{\mathrm{max}}$ and $F_{\mathrm{min}}$ are the maximum and minimum planet fluxes respectively, and $\delta$ is the hotspot offset (positive value means an eastward offset). The planet's dayside flux $F_d$ (i.e., occultation depth) and nightside flux $F_n$ are derived as the value of $F_p(\phi)$ at $\phi=\pi$ and $\phi=0$ respectively\footnote{Our code actually transforms Eq.\,(\ref{eqn:planet_flux}) as a function of $F_d$ and $F_n$ instead of  $F_{\mathrm{max}}$ and $F_{\mathrm{min}}$.}. The semi-amplitude of the atmospheric phase variation $A_{\mathrm{atm}}$ is $(F_{\mathrm{max}} -  F_{\mathrm{min}})/2$.

Following the recommendation of \citet{Akinsanmi2023OnCurves}, we account for the planetary tidal deformation in the out-of-transit phases by multiplying the planet's phase variation (Eq.\,\ref{eqn:planet_flux}) by the normalized phase-dependent projected area of the ellipsoid (Eq.\,\ref{eqn:varying_area}) to have 
\begin{equation}
    F_p(\phi)^{\mathrm{def}}= F_p(\phi)\,\frac{A(\phi)}{A(0)}.
    \label{eqn:def_Fp}
\end{equation}

\noindent This implies that the phase variation of the deformed planet depends on \hf, \Rv, and $Q_M$. \citet{bell-2019-WASP-12} employed a similar approach to model the tidal deformation of \plname in an attempt to explain its anomalous phase curve shape in the \spitzer 4.5$\mu$m band. However, the planet's shape was modeled as a biaxial ellipsoid (instead of the triaxial model used here). They found that even if tidal deformation might contribute to the phase curve, it is not sufficient to explain the observed anomaly present only in the 4.5$\mu$m phase curve. Similarly, in the analysis of the \hst and \spitzer phase curves of WASP-103\,b, \citet{kreid18} accounted for the planetary deformation by including the normalized phase-dependent projected area of the ellipsoid. However, instead of fitting the shape of the planet, it was fixed based on the tabulated predictions in \citet{Leconte2011}.
 
\subsubsection{Light travel time}

We corrected for light-travel time across the planetary system by converting the observation times ($t_{\mathrm{obs}}$) into a reference time ($t_{\mathrm{ref}}$). The reference time accounts for the projected distance between the current position of the planet along the orbit and its position at inferior conjunction. This is given for a circular orbit as \citep{deline2022}
\begin{equation}
    t_{\mathrm{ref}} = t_{\mathrm{obs}} - \frac{a}{c}\left(1-\cos{\phi}\right)\sin{i}.
\end{equation}

\subsection{Fitting process}
To obtain results regarding different planetary properties, we perform different model fits to the datasets:
\begin{itemize}
\item We analyzed all datasets by performing a global fit using the ellipsoidal and spherical planet models and comparing the results (\S\ref{sect:pc_analysis}). This analysis allows to constrain the shape and atmospheric properties of the planet.
\item We analyzed the \cheops transit observations individually (\S\ref{sect:tra_analysis}) to derive transit timings for orbital decay analysis . 
\item We analyzed the \cheops occultation observations individually (\S\ref{sect:occult_analysis}) to measure the occultation depth of each visit and probe for potential atmospheric variability.  
\end{itemize}

In these fits, the astrophysical and systematic trends are modeled simultaneously. In cases where we perform model comparison, we sample the parameter space using the nested sampling algorithm, \texttt{dynesty} \citep{speagle} which provides posteriors of the fit and also the Bayesian evidence for each model. We used 1000 live points to explore the parameter space until the estimated log-evidence was smaller than 0.1.  In other fits, we use the affine-invariant MCMC ensemble sampler implemented in \texttt{emcee} \citep{Foreman-Mackey2013}. The results of these analyses are discussed in Sections  \ref{sect:tidal_deformation}, \ref{sect:atmosphere} and \ref{sect:tidal_decay}.

\subsection{Phase curve analysis}
\label{sect:pc_analysis}
We used \texttt{dynesty} to fit the light curve model (Eq.\,\ref{eqn:pc_flux}) to all the datasets simultaneously, considering both a spherical and an ellipsoidal planet. We fit the \cheops and \tess observations using the full phase curve model, and the \spitzer observations using only the transit component. We fit for the orbital parameters: planetary period $P$, mid-transit time $T_0$, planet-star radius ratio $R_p/R_{\star}$, scaled semi-major axis $a/R_{\star}$ and impact parameter $b$. We also fit for the phase curve parameters: $F_d$, $F_n$, $\delta$, $A_{\mathrm{EV}}$, and $A_{\mathrm{DB}}$. Fitting for the total EV amplitude, $A_{\mathrm{EV}}$, ensures that we consider all possible combinations of the component limb and gravity darkening coefficients (in Eqs. \ref{eqn:A_EV} and \ref{eqn:alpha_EV}) and not just their theoretically estimated values.

As seen in Table\,\ref{tab:pc_fit}, we assumed wide uniform priors on all parameters except for the LDCs which have Gaussian priors as derived in Section\,\ref{sect:limb_darkening}, and $A_{\mathrm{DB}}$ for which we used Gaussian priors centered on theoretical values calculated from Eq.\,(\ref{eqn:alpha_db}) for each passband with a standard deviation of 0.2. The prior on $F_n$ includes negative values to ensure Gaussian posteriors and avoid overestimating the nightside flux. For the fit using the ellipsoidal planet model, we adopt a uniform prior for \hf and a normal prior for the log of the mass ratio ($\log Q_M$) based on literature values.

We performed the analysis using a two-step fitting process in which an initial fit of the data from each instrument is performed to estimate the noise properties of each visit/sector \citep[e.g.,][]{Lendl2017, wong2021_tess2ndyearPCs, Demory202355CHEOPS}. We estimated the amplitude of additional white noise $\beta_w$, which is calculated as the ratio of the residual RMS to the mean photometric uncertainty. We also calculated the amplitude of the red noise $\beta_r$ by taking the average of the ratio of the binned residuals at several timescales to the expected Gaussian $1/\sqrt{n}$ noise scaling at that timescale \citep[e.g.,][]{Winn2008TheXO3b}. We then rescaled the photometric uncertainties of the light curve by multiplying them by $\beta_w\beta_r$. This method allows us to propagate the extra noise contributions to the best-fit parameters. Fitting instead for a jitter parameter to account for extra white noise in each visit/sector of the instruments would introduce 54 additional parameters to the global fit of the datasets, making the convergence time of the fit much longer. We chose not to model the red noise using Gaussian Processes since this might be capable of absorbing the subtle signal of tidal deformation in the light curve.

\begin{figure*}[!ht]
    \centering
    \includegraphics[width=\textwidth]{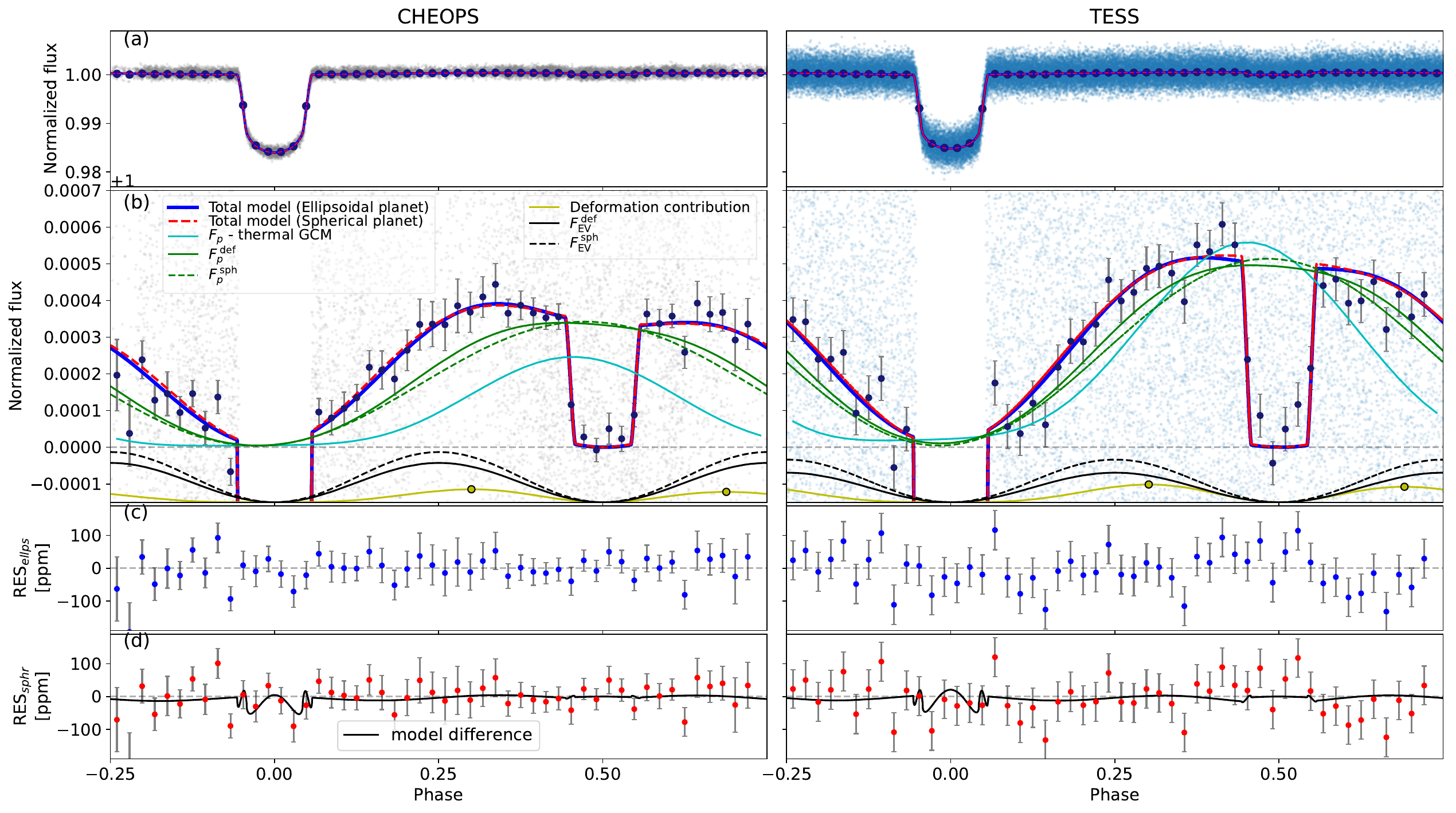}
    \caption{\cheops (left) and \tess (right) phase curves of WASP-12\,b. \textbf{\textit{Panels (a):}} The total phase curve models overplotted on the detrended and phase-folded data. \textbf{\textit{Panels (b):}} Zoom of panel (a) showing the different components of the total phase curve. The solid blue curve is the total phase curve model for an ellipsoidal planet, while the red dashed curve is for a spherical planet. The solid and dashed green curves represent the best-fit planetary phase variation component ($F_p$) for the deformed and spherical planets, respectively. The cyan curve is the computed thermal-only 3D GCM  for the planet (see \S\ref{sect:3D_GCM}) which shows for \cheops that some reflection from the atmosphere is required to reach the amplitude of the green curves. The black curves at the bottom of this panel show the best-fit stellar ellipsoidal variation ($F_{\mathrm{EV}}$) from the deformed (solid) and spherical (dashed) planet model fits. The yellow curve represents the contribution of tidal deformation to the total phase curve, with circles denoting the peaks. \textbf{\textit{Panels (c) \& (d):}} The residuals of the phase curve fits using the ellipsoidal (blue) and spherical (red) planet models. The difference between the models (ellipsoidal–spherical) representing the deformation signature in each passband is overplotted on the spherical model residuals.}
    \label{fig:PCfit}
\end{figure*}

\begin{figure*}[!ht]
    \centering
    \includegraphics[width=\linewidth]{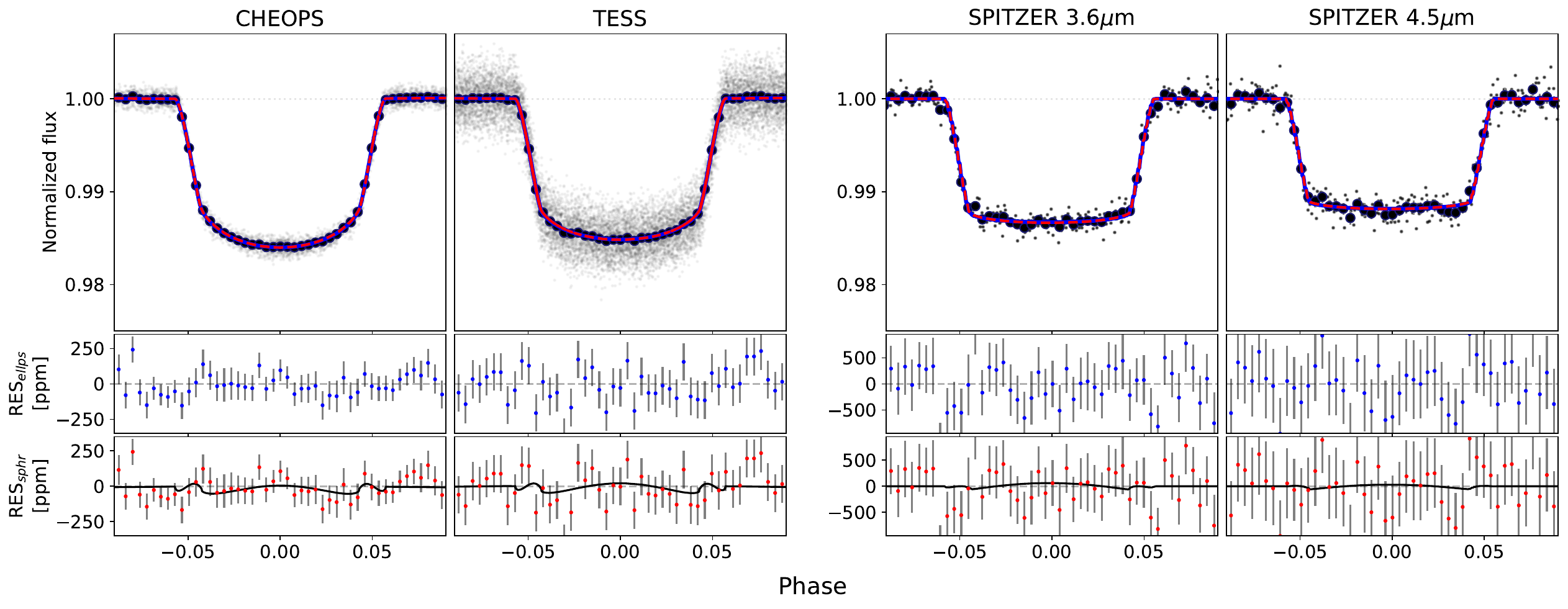}
    \caption{Phase-folded transits of \plname from the different instruments with the best-fit models overplotted. For each instrument, the 5-min binned residuals for the ellipsoidal (blue) and spherical planet (red) model fits are shown. The difference between both models is again overplotted on the spherical model residuals.}
    \label{fig:def_fit}
\end{figure*}

After determining the $\beta_w\beta_r$ for all datasets, the global results are obtained from the final joint fit of the datasets. The orbital parameters ($P$, $T_0$, $b$, $a/R_{\star}$) are common between the datasets while other passband-dependent parameters including \Rv/$R_{\star}$ are different between the datasets. We also fit the models using fixed quadratic ephemeris determined from our orbital decay result in Section\,\ref{sect:tidal_decay}, and find that the derived parameters are consistent within 1$\sigma$. The results of the spherical and ellipsoidal fits are given in Table\,\ref{tab:pc_fit} while the phase-folded data and best-fit models are shown in Fig.\,\ref{fig:PCfit} and Fig.\,\ref{fig:def_fit}. 

The results of the phase curve fit are discussed in Sections\,\ref{sect:pc_constraints} and \ref{sect:pc_atm_constraints}.

\subsection{\cheops transit analysis}
\label{sect:tra_analysis}
With the aim of deriving precise transit times, we individually analyzed the 22 \cheops transit observations using the $F_{\mathrm{tra}}$ model in Eq.\,(\ref{eqn:pc_flux}). Since the phase variation signals ($F_p$ and $F_{\star}$) impact the transit baseline of each visit, we first subtracted out the best-fit phase variation signals determined from the phase curve fit 
(in \S\,\ref{sect:pc_analysis}). We propagate the uncertainties of the $F_p$ and $F_{\star}$ models by quadratically adding their standard deviations to the flux error bars. Each transit was allowed to have a unique mid-transit time and systematics model, but the shape parameters (i.e., $R_p/R_{\star}$, $b$, $a/R_{\star}$) were constrained using Gaussian priors based on the parameter posteriors from the spherical planet phase curve fit. We sample the parameter space using \texttt{emcee}.

Figure\,\ref{fig:transit_obs} shows the best-fit transit+systematics model overplotted on each transit light curve. It also shows the best-fit transit model overplotted on each detrended light curve and annotated with the obtained timing precision. Finally, the residual of each light curve fit is shown with the measured RMS. The mid-transit times obtained for the transits and their $1\sigma$ uncertainties are given in Table\,\ref{tab:transit_times} with a mean precision of 27.5\,s.


\subsection{\cheops occultation analysis}
\label{sect:occult_analysis}
With the aim of measuring the individual occultation depths, we individually analyzed the 26 \cheops occultation observations using the $F_{\mathrm{occ}}$ model in Eq.\,(\ref{eqn:pc_flux}). Similar to the transit analyses, we also subtracted out the best-fit phase variation signals ($F_p$ and $F_{\star}$) determined from the phase curve fit and propagated their uncertainties
to the flux error bars. Here, we fit for the occultation depth but fixed the orbital and shape parameters  (i.e., $P$, $R_p/R_{\star}$, $b$, $a/R_{\star}$) to the best-fit \cheops values obtained from the phase curve fit. However, since precise timing measurements cannot be obtained from the individual occultations due to the low signal-to-noise, we use Gaussian prior on $T_0$ based on the posterior from the phase curve fit. Figure\,\ref{fig:occulation_obs} shows the best-fit occultation+systematics model overplotted on each occultation light curve. It also shows the best-fit occultation model overplotted on each detrended light curve and lastly, the residuals of each light curve fit and its RMS.

We further jointly analyzed the occultation observations within each of the three seasons, freely fitting for the mid-occultation time and occultation depth. The epoch of the occultation time for each season was set to that of the occultation closest to the center of the time series. The resulting timings for each of the three seasons (labeled S1–S3) are listed in Table\,\ref{tab:transit_times}. 

Section\,\ref{sect:variability} discusses the measured occultation depths with respect to previously reported hints of atmospheric variability in dayside of \plname. 

\begin{table*}[th!]
\centering
\caption{Result of the fit to the \cheops and \tess phase curves. We note that $\mathcal{N}(\mu,\sigma)$ represents a Gaussian prior with mean $\mu$ and standard deviation $\sigma$ while $\mathcal{U}(a,b)$ is a uniform prior between $a$ and $b$.}
\label{tab:pc_fit}
\begin{tabular}{lllll}
\hline
Parameter &
  Symbol &
  \multicolumn{1}{l|}{Prior} &
  \multicolumn{2}{c}{Posterior} \\ \cline{4-5} 
 &
   &
  \multicolumn{1}{l|}{} &
  \begin{tabular}[c]{@{}l@{}}Spherical planet\end{tabular} &
  \begin{tabular}[c]{@{}l@{}}Ellipsoidal planet\end{tabular} \\ \hline
  
Orbital Period\,[d] &
  $P$ &
  \multicolumn{1}{l|}{$\mathcal{N}$(1.0914,0.0001)} &
  ${1.0914185}\pm2$E-7 &
  ${1.0914185}\pm2$E-7 \\[0.2cm]
  
Transit time [BJD$_\mathrm{TBD}$–2459000]&
  $T_0\,^\diamond$ &
  \multicolumn{1}{l|}{$\mathcal{U}$(196,197)} &
  $196.624122\pm6.6$E-4 &
  $196.624121\pm6.6$E-4 \\[0.2cm]
  
Impact parameter &
  $b$ &
  \multicolumn{1}{l|}{$\mathcal{U}$(0,1)} &
  $0.392\pm{0.011}$ &
  $0.369\pm{0.016}$ \\[0.2cm]
  
Scaled semi-major axis &
  $a/R_{\star}$ &
  \multicolumn{1}{l|}{$\mathcal{U}$(2.5, 3.5)} &
  ${3.006}\pm{0.013}$ &
  ${3.036}\pm{0.018}$ \\[0.2cm]
  
Love number &
  $h_2$ &
  \multicolumn{1}{l|}{$\mathcal{U}$(0,2.5)} &
  \multicolumn{1}{c}{–} &
  $1.55_{-0.49}^{+0.45}$ \\[0.2cm]
  
Log planet-to-star mass ratio &
  $\log Q_M$ &
  \multicolumn{1}{l|}{$\mathcal{N}$(-6.935, 0.025)} &
  \multicolumn{1}{c}{–} &
  $-6.931\pm{0.022}$ \\[0.3cm]

\begin{tabular}[c]{@{}l@{}}Planet–star radius ratio$^{\ast}$\\ \\ \\ \\\end{tabular}&
\begin{tabular}[c]{@{}l@{}}\Rv/$R_{\star_{_{\mathrm{CHEOPS}}}}$\\
                           \Rv/$R_{\star_{_{\mathrm{TESS}}}}$\\
                           \Rv/$R_{\star_{_{\mathrm{Spitzer3.6}}}}$\\
                           \Rv/$R_{\star_{_{\mathrm{Spitzer4.5}}}}$\\\end{tabular} &
\multicolumn{1}{l|}{\begin{tabular}[c]{@{}l@{}}$\mathcal{U}$(0, 0.25)\\ \\ \\ \\ \end{tabular} } &
\begin{tabular}[c]{@{}l@{}} $0.1177\pm0.0002$\\   
                            $0.1164\pm0.0002$\\
                            $0.1133\pm0.0004$\\  
                            $0.1071\pm0.0006$\end{tabular} & 
\begin{tabular}[c]{@{}l@{}} $0.1242\pm0.0023$\\
                            $0.1226\pm0.0023$ \\
                            $0.1188\pm0.0020$\\ 
                            $0.1115\pm0.0017$
                            \end{tabular} \\[0.5cm]

EV semi-amplitude\,[ppm] &
  \begin{tabular}[c]{@{}l@{}}$A_{\mathrm{EV}_{_{\mathrm{CHEOPS}}}}$\\
                            $A_{\mathrm{EV}_{_{\mathrm{TESS}}}}$ \end{tabular}&
  \multicolumn{1}{l|}{$\mathcal{U}$(0,200)} &
  \begin{tabular}[c]{@{}l@{}}$69\pm14$\\ 
                             $58\pm14$
                             \end{tabular} &
  \begin{tabular}[c]{@{}l@{}}$54\pm15$\\ 
                             $40\pm15$
                             \end{tabular} 
                             \\[0.5cm]
  
DB semi-amplitude\,[ppm] &
  \begin{tabular}[c]{@{}l@{}}$A_{\mathrm{DB}_{_{\mathrm{CHEOPS}}}}$\\
                            $A_{\mathrm{DB}_{_{\mathrm{TESS}}}}$ \end{tabular}&
  \multicolumn{1}{l|}{\begin{tabular}[c]{@{}l@{}}$\mathcal{N}$(2.92, 0.2)\\                                                                $\mathcal{U}$(2.32, 0.2)
                                                 \end{tabular}} &
  \begin{tabular}[c]{@{}l@{}}$2.92\pm0.17$\\ 
                             $2.31\pm0.17$
                             \end{tabular} &
  \begin{tabular}[c]{@{}l@{}}$2.91\pm0.17$\\ 
                             $2.31\pm0.17$
                             \end{tabular} \\[0.5cm]
  
\begin{tabular}[c]{@{}l@{}}Dayside flux\,[ppm]\\ (occultation depth)\end{tabular} &
  \begin{tabular}[c]{@{}l@{}}$F_d/F_{\star_{_{\mathrm{CHEOPS}}}}$\\
                             $F_d/F_{\star_{_{\mathrm{TESS}}}}$\\\end{tabular}&
  \multicolumn{1}{l|}{$\mathcal{U}$(0, 1000)} &
  \begin{tabular}[c]{@{}l@{}}$340\pm24$\\ 
                             $513\pm29$
                             \end{tabular} &
  \begin{tabular}[c]{@{}l@{}}$333\pm24$\\ 
                             $493\pm29$
                             \end{tabular} \\[0.5cm]
  
Nightside flux\,[ppm] &
  \begin{tabular}[c]{@{}l@{}}$F_n/F_{\star_{_{\mathrm{CHEOPS}}}}$\\
                         $F_n/F_{\star_{_{\mathrm{TESS}}}}$\\\end{tabular}&
  \multicolumn{1}{l|}{$\mathcal{U}$(-100, 100)} &
  \begin{tabular}[c]{@{}l@{}}$6.2\pm23$\\ 
                             $6.1\pm21$
                             \end{tabular} &
  \begin{tabular}[c]{@{}l@{}}$7.0\pm23$\\ 
                             $11.8\pm21$
                             \end{tabular} \\[0.5cm]
  
Hotspot offset\,$[\degree]$ &
  \begin{tabular}[c]{@{}l@{}}$\delta_{_{\mathrm{CHEOPS}}}$\\
                             $\delta_{_{\mathrm{TESS}}}$\\\end{tabular}&
  \multicolumn{1}{l|}{$\mathcal{U}$(-90,90)} &
  \begin{tabular}[c]{@{}l@{}}$9.8\pm4.5$\\ 
                             $6.3\pm2.7$
                             \end{tabular} &
  \begin{tabular}[c]{@{}l@{}}$9.3\pm4.6$\\ 
                             $5.8\pm2.6$
                             \end{tabular} \\ \hline
  
\multicolumn{5}{l}{\textbf{Derived parameters}}\\ \hline

\begin{tabular}[c]{@{}l@{}}\textit{Effective} planetary radius ratio\\
                            mid-occultation\end{tabular} &
\begin{tabular}[c]{@{}l@{}} $R_p^{\mathrm{eff}}/R_{\star_{_{\mathrm{CHEOPS}}}}$\\   
                    $R_p^{\mathrm{eff}}/R_{\star_{_{\mathrm{TESS}}}}$\end{tabular} & 
  \multicolumn{1}{c|}{–} &
  \begin{tabular}[c]{@{}l@{}}$0.11770\pm{0.00020}$\\
                             $0.11640\pm{0.00020}$
                             \end{tabular} &
  \begin{tabular}[c]{@{}l@{}}$0.11730\pm0.00027$\\               
                             $0.11600\pm0.00025$
                             \end{tabular} \\[0.5cm]

\begin{tabular}[c]{@{}l@{}}Ellipsoid axis ratios\\ \\ \\ \\\end{tabular}&
\begin{tabular}[c]{@{}l@{}}$r_1$\,:\,$r_2$\,:\,$r_3\,_{_{\mathrm{CHEOPS}}}$\\
                            $r_1$\,:\,$r_2$\,:\,$r_3\,_{_{\mathrm{TESS}}}$\\
                            $r_1$\,:\,$r_2$\,:\,$r_3\,_{_{\mathrm{Spitzer3.6}}}$\\
                            $r_1$\,:\,$r_2$\,:\,$r_3\,_{_{\mathrm{Spitzer4.5}}}$\\
                            \end{tabular}  &
  \multicolumn{1}{c|}{\begin{tabular}[c]{@{}l@{}}–\\– \\– \\– \\\end{tabular} } &

  \multicolumn{1}{c}{\begin{tabular}[c]{@{}l@{}} –\\   –\\–\\  –\end{tabular}} &
\begin{tabular}[c]{@{}l@{}} 1.23\,:\,1.06\,:\,1.0  \\
                            1.22\,:\,1.05\,:\,1.0 \\
                            1.20\,:\,1.05\,:\,1.0\\ 
                            1.16\,:\,1.04\,:\,1.0\end{tabular} \\[0.4em]

\hline
\begin{tabular}[c]{@{}l@{}}Planet phase variation\\ semi-amplitude\,[ppm] \end{tabular} &
  \begin{tabular}[c]{@{}l@{}}$A_{\mathrm{atm}_{_{\mathrm{CHEOPS}}}}$\\
                             $A_{\mathrm{atm}_{_{\mathrm{TESS}}}}$\\\end{tabular}&
  \multicolumn{1}{c|}{–} &
  \begin{tabular}[c]{@{}l@{}}$170\pm16$\\           
                             $254\pm14$
                             \end{tabular} &
  \begin{tabular}[c]{@{}l@{}}$165\pm16$\\   
                             $241\pm14$
                             \end{tabular} \\[0.5cm]

\begin{tabular}[c]{@{}l@{}}Planetary mass\\ from EV\,[$M_{\mathrm{Jup}}$] \end{tabular} &
  \begin{tabular}[c]{@{}l@{}}$M_{p_{_{\mathrm{CHEOPS}}}}$\\
                             $M_{p_{_{\mathrm{TESS}}}}$\\\end{tabular}&
  \multicolumn{1}{c|}{–} &
  \begin{tabular}[c]{@{}l@{}}$2.34\pm0.48$\\
                             $2.09\pm0.49$
                             \end{tabular} &
  \begin{tabular}[c]{@{}l@{}}$1.89\pm0.53$\\
                             $1.49\pm0.58$
                             \end{tabular} \\[0.5cm]

Geometric albedo &
  \begin{tabular}[c]{@{}l@{}}$A_{g_{_{\mathrm{CHEOPS}}}}$\\
                             $A_{g_{_{\mathrm{TESS}}}}$\\\end{tabular}&
  \multicolumn{1}{c|}{–} &
  \begin{tabular}[c]{@{}l@{}}$0.089\pm0.017$\\            
                             $0.022\pm0.023$
                             \end{tabular} &
  \begin{tabular}[c]{@{}l@{}}$0.086\pm0.017$\\        
                             $0.010\pm0.023$
                             \end{tabular} \\[0.5cm]
  
Dayside temperature\,[K] &
  \begin{tabular}[c]{@{}l@{}}$T_{\mathrm{day}_{_{\mathrm{CHEOPS}}}}$\\
                             $T_{\mathrm{day}_{_{\mathrm{TESS}}}}$\\\end{tabular}&
  \multicolumn{1}{c|}{–} &
  \begin{tabular}[c]{@{}l@{}}
                            $2821\pm25$
                            \\$2915\pm25$\end{tabular} 
                            &
  \begin{tabular}[c]{@{}l@{}}
                            $2821\pm25$
                            \\$2915\pm25$\end{tabular}
                              \\ [0.5cm]

Nightside temperature\,[K]  &
  \begin{tabular}[c]{@{}l@{}}$T_{\mathrm{night}_{_{\mathrm{CHEOPS}}}}$\\
                             $T_{\mathrm{night}_{_{\mathrm{TESS}}}}$\\\end{tabular}&
  \multicolumn{1}{c|}{–} &
  \begin{tabular}[c]{@{}l@{}}
                            $<2025~(2\sigma)$
                            \\[0.15cm] 
                            $<1890~(2\sigma)$\end{tabular}
                             &
  \begin{tabular}[c]{@{}l@{}}
                            $<2126~(2\sigma)$
                            \\[0.15cm] 
                            $<1921~(2\sigma)$\end{tabular} 
                             \\ [0.5cm]

  
\hline
  
\\ \multicolumn{5}{l}{$^{\ast}$  For the spherical planet, this is equal to the transit radius.}\\ 
  
\end{tabular}
\end{table*}
\section{Tidal deformation}
\label{sect:tidal_deformation}

\subsection{Measuring deformation from phase curve}
\label{sect:pc_constraints}
From the joint fit of the \cheops, \tess, and \spitzer datasets, we compare the results of the ellipsoidal and spherical planet models. Table\,\ref{tab:pc_fit} reports the median posterior and 1$\sigma$ uncertainties of the parameters of both models. The posterior probability distributions of some relevant parameters are also shown in Fig.\,\ref{fig:def_corner} where we see greater uncertainties in the determination of the planet-star radius ratios for the ellipsoidal model due to strong correlations with the Love number.   

From the ellipsoidal planet phase curve fit, we measured a Love number of $1.55^{+0.45}_{-0.49}$ corresponding to a 3.16$\sigma$ detection. This is the first measurement of the Love number of a planet from the analysis of its full-orbit phase curve. Previous work by \citet{Barros2022} obtained a 3$\sigma$ measurement of the Love number of WASP-103\,b from the analysis of transit-only observations from different space telescopes. Since planetary tidal deformation signal in the phase curve is correlated with the stellar ellipsoidal variation signal, we additionally follow the strategy of \citet{Barros2022} to measure the deformation in the transit-only regions where the stellar ellipsoidal variation is insignificant. We measured \hf=$1.56^{+0.47}_{-0.52}$ in agreement with the phase curve derived value, but at a slightly reduced significance of $\sim$3$\sigma$. This indicates that the detection of deformation is robust against the modeling of ellipsoidal variation and that the inclusion of out-of-transit data in our phase curve fit slightly enhances the detection of deformation \citep[see][]{Akinsanmi2023OnCurves}. Both detections of tidal deformation from its induced subtle effects on light curves have been facilitated by \cheops, allowing to reach the 3$\sigma$ measurement significance. We further assess the significance of our detection by computing the Bayes factor from the log-evidence of each model obtained from the \texttt{dynesty} fit. The Bayes factor, $B_{\mathrm{ES}}$ is computed as the exponent of the difference in log-evidence between the ellipsoidal and spherical models. We obtained a Bayes factor of 6.7 in positive favor of the ellipsoidal planet model \citep{kass}.

Figure\,\ref{fig:PCfit} shows the best-fit phase curve models to the \cheops and \tess light curves and the contribution of the component signals. We see that the major impact of tidal deformation on the total phase curve is the modification of the planet's atmospheric phase variation $F_p$ (green curves). Compared to the spherical planet case, $F_p$ for the deformed planet features additional flux contribution between transit and eclipse due to the larger and varying projected size of the ellipsoid at these phases. The contribution of tidal deformation to the total ellipsoidal planet phase curve (yellow curve at the bottom of panel \textit{b}) peaks between quadrature and eclipse (at phases 0.3 and 0.69). The spherical planet model fit attempts to account for the deformation contribution by increasing the amplitude of the stellar ellipsoidal variation $F_{\mathrm{EV}}^{sph}$ (dashed black curve). However, since the peaks of $F_{\mathrm{EV}}$ occur at quadratures and are always spaced by 0.5 phases, it is unable to completely absorb the deformation signal (which has shorter peak spacing) even if it is shifted in phase. The bottom two panels of Fig.\,\ref{fig:PCfit} show the residuals from the fits. In each passband, the difference between the ellipsoidal and spherical planet models is overplotted on the spherical planet fit residuals, highlighting the deformation signature – that is, the remaining signal due to deformation that cannot be accounted for by a spherical planet model. The deformation signature is concentrated within the in-transit phases while the out-of-transit phases show only slight variations. This is due to the low amplitude of the planetary phase variation in these optical bands (a few hundred ppm) which influences the out-of-transit deformation signature. Phase curve observations in the infrared, where the amplitude of the planet's atmospheric phase variation can reach thousands of ppm, will lead to a significant increase in the deformation signature outside transit, thereby allowing a more precise detection of the deformation and Love number measurement \citep{Akinsanmi2023OnCurves}.

Figure\,\ref{fig:def_fit} shows the transit phases of the fit for all the datasets. We see that the amplitude and shape of the deformation signature are passband dependent, respectively, due to the varying size of the planet and also different limb darkening compensation at each passband. The ratios of the ellipsoidal planet axes are also passband-dependent since the asymmetry between the axes depends on the radius at that passband (Eq.\,\ref{eqn:asym_par}).  Using Eqs.\,(\ref{eqn:ell_axis}) and (\ref{eqn:asym_par}) with the derived values from the ellipsoidal planet model fit, we calculate the axis ratios, $r_1$:$r_2$:$r_3$, in the different passbands and report them in Table\,\ref{tab:pc_fit}. As opposed to the large $r_1$/$r_3$ ratio of 1.5 estimated in \citep{cowan2012-wasp-12} to explain the \spitzer 4.5$\mu$m phase curve anomaly, our fit to the datasets results in a more physical ratio of 1.16 in this passband. Therefore, the source of the \spitzer 4.5$\mu$m remains unexplained and requires further investigation.

\begin{figure}
\centering
    \includegraphics[width=\linewidth]{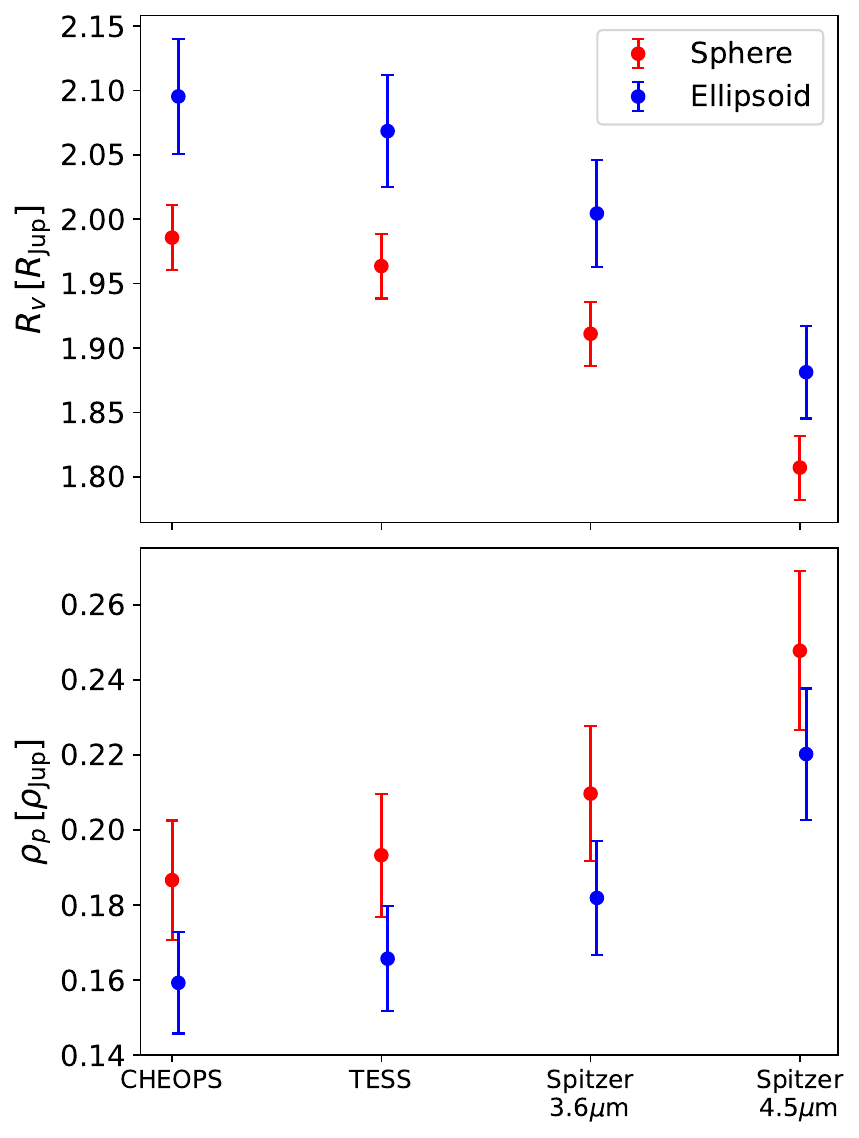}
    \captionof{figure}{Physical radius and density for \plname derived from the spherical and ellipsoidal planet model fits.}
    \label{fig:rad_dens}
\end{figure}

Comparing the posterior parameters between the ellipsoidal and spherical planet model fits, we find that most parameters differ only by $\lesssim$2$\sigma$ due to the limited precision of \hf obtained from these datasets. Figure\,\ref{fig:rad_dens} compares the physical radius and mean density of the planet in the different passbands, calculated using the posteriors from the fits and stellar parameters in Table\,\ref{tab:stellar_table}. The results show that the assumption of sphericity leads to an underestimation of the volumetric radius of \plname by 4.0–5.2\% and an overestimation of the planetary density by $\gtrsim$12\%, in line with theoretical expectations \citep{Leconte2011,burton,correia14}. The biases in radius and density are respectively only $\sim$2$\sigma$ and $\sim$1$\sigma$ significant due to the limited precision of the derived \hf from these datasets and the planetary mass from radial velocity (RV). Therefore, an increase in photometric and RV precisions will lead to more significant parameter biases and the derived properties of the planet if sphericity is assumed \citep{Berardo2022TidalCompositions}.

For both deformed and spherical model phase curve fits, the measured stellar ellipsoidal variation semi-amplitude $A_{\mathrm{EV}}$ is larger in the \cheops band compared to \tess. This is due to the passband-dependent limb and gravity darkening parameters in $\alpha_{\mathrm{EV}}$ in Eq.\,(\ref{eqn:alpha_EV}).  The measured $A_{\mathrm{EV}}$ in both bands can be used to independently estimate the mass of \plname using Eqs.\,(\ref{eqn:A_EV}) and (\ref{eqn:alpha_EV}), and we expect to obtain the same mass estimate in both passbands. First, the linear limb-darkening parameter for each passband was estimated using \texttt{LDCU}, while the gravity darkening coefficient was estimated from the tables in \cite{Claret2021LimbCHEOPS}  and \cite{Claret2017LimbVelocities} for the \cheops and \tess bands, respectively. We use the stellar mass from Table\,\ref{tab:stellar_table} and orbital parameters from the joint fit in Table\,\ref{tab:pc_fit}. For the spherical planet model, we derive a planetary mass of $2.34\pm0.48$\,$M_{\mathrm{Jup}}$ from \cheops parameters and $2.09\pm0.49$\,$M_{\mathrm{Jup}}$ from \tess which are consistent with one another within $1\sigma$. These masses are higher but consistent with the value of $1.47\pm0.073$\,$M_{\mathrm{Jup}}$ derived from radial velocity \citep[RV;][]{Collins2017TRANSITPLANETS} at 1.2–1.8$\sigma$. For the ellipsoidal planet model, we derive $1.89\pm0.53$\,$M_{\mathrm{Jup}}$ for \cheops and $1.49\pm0.58$\,$M_{\mathrm{Jup}}$ for \tess which are more consistent with the RV value at $<$0.8$\sigma$. Although the discrepancy between the masses from EV and RV in the spherical model case is not significant due to the large uncertainties on EV masses, we see a slight indication that accounting for deformation reconciles both estimates better.

The derived \hf for \plname is unexpectedly close to the value of Jupiter despite its higher insolation and mass. This was also the case for WASP-103\,b with the measured \hf of 1.585 \citep{Barros2022}. Theoretical models predict a decrease in the Love number with mass above $\sim$1\,$M_{\mathrm{Jup}}$ since more massive objects are more compressible and therefore tend to become more centrally condensed \citep{Leconte2011}. Lower \hf values are also expected at higher equilibrium temperatures due to the lower density of the planetary envelope compared to the core \citep{kramm12, Wahl2021TidalJupiters}. Nonetheless, our derived \hf value is consistent with the theoretical models \citep[e.g.,][]{Leconte2011, Wahl2021TidalJupiters} predicting a maximum \hf of 1.6 for tidally locked hot Jupiters. It has been suggested that the ellipsoidal planet shape model might not sufficiently account for nonlinear tidal response of the planet causing it to systematically overestimate the Love number \citep{Wahl2021TidalJupiters}. However, transit models that account for these are not currently available and will require more parameters to model the planetary shape which will be strongly correlated.

Following the procedure of \citet{Buhler2016}, we attempt to estimate the core mass fraction of the \plname using the measured $h_2$ value. We found that the 3$\sigma$ measurement of the Love number does not provide valuable constraints as the result remain consistent with a core mass fraction of 0  and 1. Indeed, \citet{Akinsanmi2023OnCurves} showed that valuable constraints on the core mass fraction require $h_2$ precisions higher than 4$\sigma$. They also showed that a single \texttt{JWST} phase curve of a target such as \plname is capable of attaining 17$\sigma$ measurement of $h_2$ due to the large phase curve amplitude in the infrared, the reduced effect of limb darkening, and the unparalled precision of the instrument.

\subsection{Impact of limb darkening }
\label{sect:ldc_impact}

As most of the optical band deformation signature is concentrated within the transit phases, the detection is sensitive to the modeling of the limb darkening profile. In the fit for tidal deformation, we used the method described in Section\,\ref{sect:limb_darkening} to derive priors on the LDCs based on model intensity profiles from the two spectral synthesis libraries (we call this our fiducial analysis). We also explore two alternative methods to leveraging the generated model intensity profiles, which involve simultaneously fitting the model intensity profiles and the transit observations in order to find the best-fit LD profile.\\

\noindent \textbullet \, Alternative-1: This approach merges the \textsc{PHOENIX} and \textsc{ATLAS} model intensity profiles to create a new joint intensity profile whose 1$\sigma$ uncertainty at each $\mu$ encompasses the 1$\sigma$ uncertainty of the individual model profiles (see Fig.\,\ref{fig:ldc_plot}b).  The joint intensity profile is fitted with the power-2 LD law simultaneously with the transit observations so that a joint likelihood of the suggested limb darkening profile and transit model is obtained (for each passband) at each iteration in our sampling process. This approach allows the joint model intensity profile and transit observations to determine the best-fit limb darkening profile. The parameters of the LD law here have uninformative wide priors since their values are constrained during the joint fit.

\noindent\textbullet \, Alternative-2: In this approach, a uniform bound is defined that spans the median of both model intensity profiles (see Fig.\,\ref{fig:ldc_plot}c). During the likelihood estimation, each $\mu$ point in the suggested limb darkening profile that is anywhere within the defined bounds is assigned zero likelihood. Profile points outside the bound are still acceptable but have lower likelihood values (Gaussian) depending on their distance from the bounds. This approach is agnostic to the eventual choice of limb darkening profile as long as its points are within or close to the bounds defined by both theoretical stellar intensity profiles.

We perform the ellipsoidal planet model fits to the data again using the aforementioned limb darkening alternative approaches and compare the results with our fiducial analysis. The result is given in Table\,\ref{tab:ld_comparison} where a consistent value of $h_2$ is derived with the 3 methods, indicating that our fit is robust against the modeling of limb darkening.

\begin{table}[ht]
\centering
\caption{Sensitivity of derived Love number on different approaches of modeling limb darkening.}
\label{tab:ld_comparison}
\begin{tabular}{lcc}
\hline\hline
LD method     & $h_f$                        & Significance  \\ \hline
Fiducial      & $1.55^{+0.45}_{-0.49}$    & 3.16$\sigma$   \\[0.5em]
Alternative–1 & $1.53^{+0.45}_{-0.51}$    & 3.00$\sigma$   \\[0.5em]
Alternative–2 & $1.54^{+0.47}_{-0.52}$    & 2.96$\sigma$   \\ 
\hline
\end{tabular}
\end{table}


\section{Atmospheric characterization}
\label{sect:atmosphere}

Phase curves provide a wealth of information to characterize the atmosphere of a planet such as the day and nightside temperatures, the longitudinal temperature and reflectivity map, and also the efficiency of heat transport among others \citep[see e.g.,][]{Cowan_Agol_2011, Shporer2017-PCreview,Parmentier2018}. In this section, we infer the atmospheric properties of \plname from our \cheops and \tess phase curve analyses (\S \ref{sect:pc_analysis}). Unless otherwise stated, the discussion below is based mostly on the phase curve fit using the ellipsoidal planet model. However, we still report the derived parameters for both planet models in  Table\,\ref{tab:pc_fit} and we found them to be in agreement within 1$\sigma$.

\subsection{Phase curve constraints}
\label{sect:pc_atm_constraints}
The results of the phase curve fit for the ellipsoidal and spherical planet model fits (Table\,\ref{tab:pc_fit}) reveal a larger planet-to-star radius ratio in the \cheops band compared to the \tess band, indicating stronger atmospheric opacity in the bluer \cheops band. We also measure a significantly lower dayside flux (occultation depth) in \cheops compared to \tess. The nightside fluxes in both passbands are consistent with zero at $<$1$\sigma$ ($2\sigma$ upper limits of $\sim$50\,ppm).  The best-fit phase curves show only marginally significant eastward phase offsets of $9\pm5\degree$ and $6\pm3\degree$ in the \cheops and \tess bands, respectively. The low nightside flux and small phase offset in both bands indicate low-efficiency day-night heat redistribution at the atmospheric layers probed by the instruments. This is consistent with the theoretical and observed trend of decreasing phase offset with increasing temperature for ultra-hot Jupiters \citep{Parmentier2016TRANSITIONSJUPITERS,Komacek2016AtmosphericDifferences, Komacek2017AtmosphericObservations} possibly due to their short radiative timescales \citep{Perna2012THEDISSIPATION}.

\subsection{Atmospheric modeling}
We model the atmosphere of \plname by performing 1D retrievals on the emission spectrum and computing the forward global circulation model. This will facilitate the proper interpretation of the phase curve, thereby enabling the determination of the relative contribution of both reflection and thermal emission to the observed occultation depths.

\begin{figure*}[!ht]
    \centering
    \includegraphics[width=0.73\linewidth]{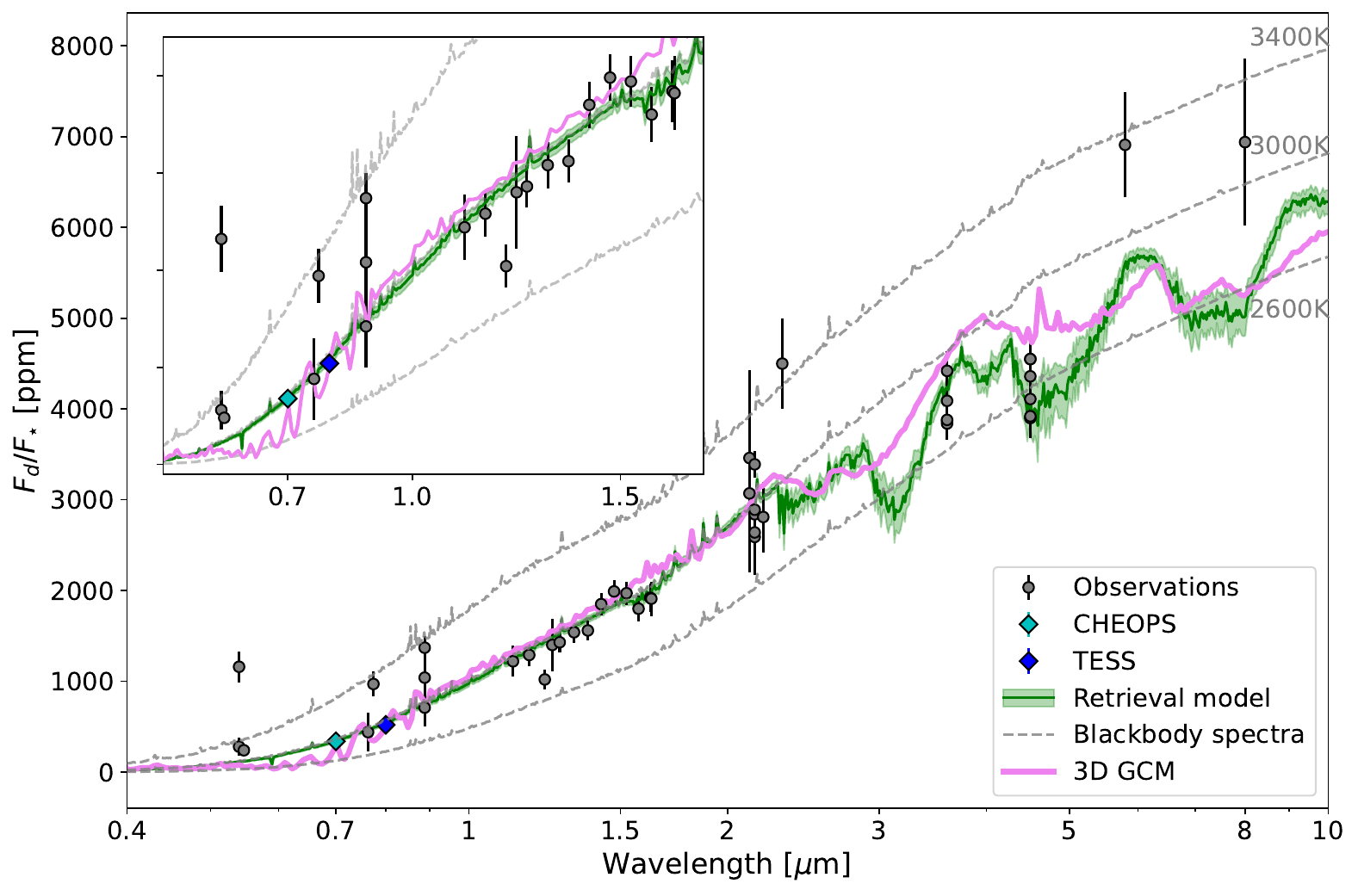}
    \includegraphics[width=0.265\linewidth]{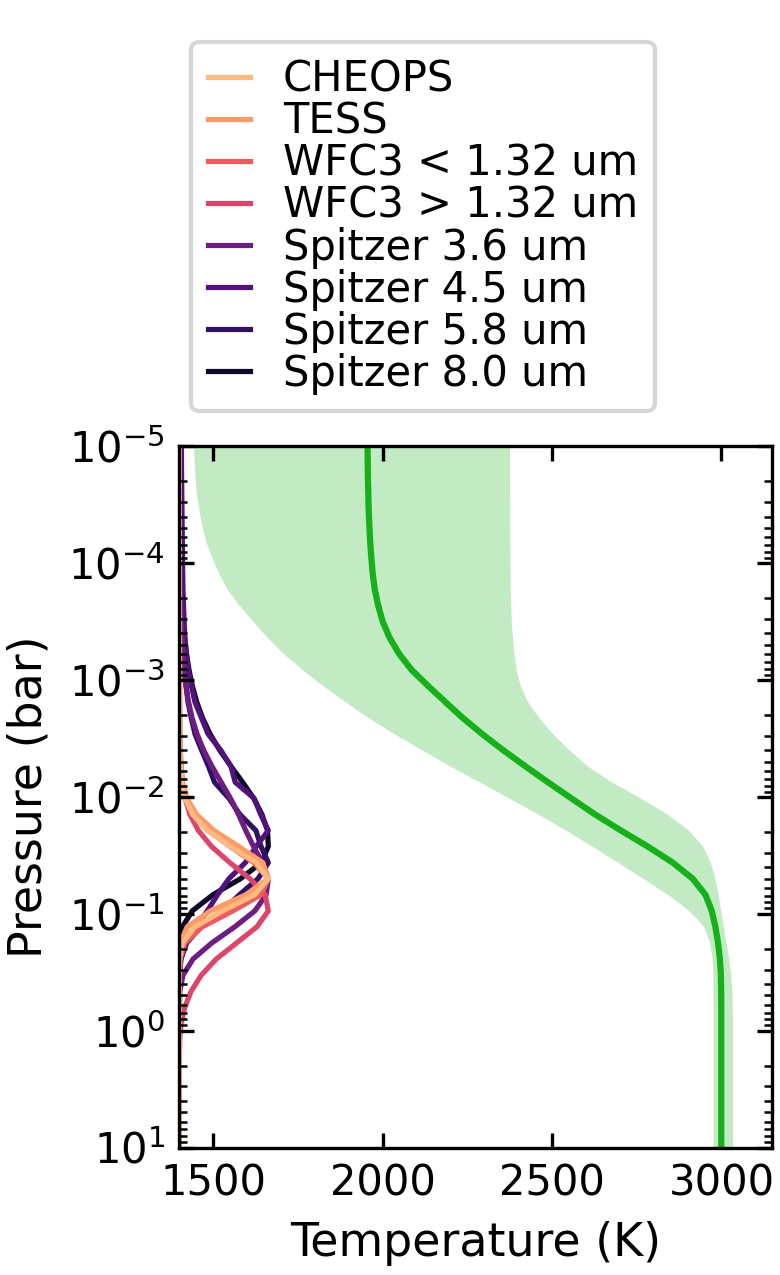}
    \caption{Eclipse spectrum of WASP-12\,b. Left: Observed occultation depths of WASP-12\,b (gray points) with the \cheops and \tess occultation depths are shown as colored diamonds.  The best-fit thermal retrieval model and its 1$\sigma$ uncertainties are overplotted in green while the forward thermal 3D GCM spectrum is plotted in violet. The gray dashed curves show the blackbody model spectra for brightness temperatures of 2600\,K, 3000\,K, and 3400\,K. Right: T-P profile for the retrieval showing a gradual temperature decrease with decreasing pressure depth between 10$^{-1}$ – $10^{-4}$\,bar indicative of no thermal inversion in the atmosphere of \plname. The contribution function at each passband is also shown indicating that \tess and \cheops essentially probe the same pressure levels within the atmosphere.} 
    \label{fig:em_spec}
\end{figure*}

\subsubsection{1D retrieval }

We model the emission spectra of \plname using the open-source {\pyratbay} framework \citep{Cubillos2021TheSpectra} to constrain its atmospheric properties based on the tabulated occultation depth measurements in \citet[][and references therein]{Hooton2019StormsWASP-12b}.  This dataset consists of several ground-based, \texttt{HST/WFC3}, and \texttt{Spitzer} measurements spanning 0.5–8$\mu$m.  Since reflection is not accounted for in {\pyratbay}, our model only considered the infrared observation ($\lambda>1.0$ $\mu$m) to avoid interference from the reflected flux at shorter wavelengths. The retrieved thermal emission spectrum can then be computed including shorter wavelengths to estimate the thermal contribution in the \cheops and \tess passbands.

We model the atmosphere of \plname between 10$^{2}$--10$^{-9}$\,bar adopting the parametric temperature--pressure (T-P) prescription of \citet{Guillot2010OnAtmospheres}. The parametric model depends on the irradiation temperature $T_{\mathrm{irr}}$, the mean thermal opacity $\log \kappa'$, and the ratio of the visible to thermal opacities $\log \gamma$. For this analysis, we modeled the composition under thermochemical equilibrium, considering the most relevant neutral and ionic species expected for hot-Jupiter atmospheres.  The chemistry was parameterized by the abundance of carbon [C/H], oxygen [O/H], and all other metals [M/H] relative to solar-abundance values.  The radiative transfer calculation considered HITEMP and ExoMol opacity line lists for H$_2$O, CO, CO$_2$, CH$_4$, HCN, NH$_3$, TiO, VO \citep{RothmanEtal2010jqsrtHITEMP, LiEtal2015apjsCOlineList, HargreavesEtal2020apjsHitempCH4, HarrisEtal2006mnrasHCNlineList, HarrisEtal2008mnrasExomolHCN, PolyanskyEtal2018mnrasPOKAZATELexomolH2O, ColesEtal2019mnrasNH3coyuteExomol, Yurchenko2015jqsrtBYTe15exomolNH3, McKemmishEtal2016mnrasVOMYTexomolVO, McKemmishEtal2019mnrasTOTOexomolTiO}, which were preprocessed using the \textsc{repack} algorithm to extract the dominant line transitions \citep{Cubillos2017apjRepack}.  Additional opacities include the Na and K resonant lines \citep{BurrowsEtal2000apjBDspectra}, collision-induced absorption from H$_2$–H$_2$ \citep{BorysowEtal2001jqsrtH2H2highT, Borysow2002jqsrtH2H2lowT}, Rayleigh opacity from H$_2$, H, and He \citep{Kurucz1970saorsAtlas}, and H$^{-}$ free-free and bound-free opacity \citep{John1988aaHydrogenIonOpacity}.  For the stellar spectrum, we used a synthetic \textsc{PHOENIX} spectrum \citep{husser} according to the stellar properties (Table\,\ref{tab:stellar_table}).  Finally, the atmospheric Bayesian retrieval employed a differential-evolution MCMC algorithm implemented in \citet{CubillosEtal2017apjRednoise} to construct posterior distributions for the atmospheric parameters.

We tested four retrieval scenarios: with and without the \texttt{Spitzer} 5.8 and 8.0$\mu$m data points; and also with and without TiO/VO absorption (since these species may or may not be present in the atmosphere).  We found statistically consistent results between all scenarios, thus in the following, we report the results of the retrieval including TiO/VO and all \texttt{Spitzer} observations.

Figure\,\ref{fig:em_spec} shows the retrieved T-P profile and thermal spectrum, which we extended over the \cheops and \tess passbands. The retrieval spectral fit is mostly driven by the  HST/WFC3 and the Spitzer 3.6 and 4.5~$\mu$m observations. The model fits most of the occultation depths relatively well, although several mutually inconsistent measurements lead to a reduced $\chi^2$ of 2.9.  The observations at wavelengths $\lambda < 2\,\mu$m are well fit by a $\sim$3000~K blackbody model, indicating that these bands probe a nearly isothermal region of the atmosphere. The depths at the \texttt{Spitzer} 3.6~$\mu$m and 4.5~$\mu$m bands imply lower brightness temperatures ($\sim$2600\,K) than in the HST/WFC3 band, possibly due to absorption by carbon-bearing species. These measurements drive the retrieved model towards a non-inverted T-P profile since these bands probe the upper atmosphere (see contribution function and T–P profile in the right panel of Fig.\,\ref{fig:em_spec}). Previous work by \citet{Stevenson2014DECIPHERINGEMISSION} and \citet{Oreshenko2017RetrievalHistory} similarly finds a noninverted T-P profile. The retrieval model underestimates the \texttt{Spitzer} 5.8~$\mu$m and 8.0~$\mu$m which would require higher brightness temperatures of $\sim$3400\,K, and thus are inconsistent with the rest of the observations. These \spitzer points have larger uncertainties and are therefore not very constraining in the fit.  

In all scenarios, we find insignificant contributions from N-bearing species, which are not expected to be prominent at high-temperature atmospheres in thermochemical equilibrium. Likewise, TiO and VO do not seem to have an impact on the retrieved spectrum, probably because the bluer end of the spectrum (where these molecules absorb the most) is nearly isothermal, which decreases the amplitude of spectral features.  On the other hand, C- and O-bearing species (CO, CO2, CH4, H$_2$O) are more prominent.  Figure~\ref{fig:pyrat_posterior} shows the retrieval posterior distribution, we find super-solar abundances for both carbon and oxygen while keeping mostly C/O > 1
ratios. For other metals, we find sub-solar abundances. These results are consistent with previous retrieval analyses of the {\plname} emission spectrum \citep[e.g.,][]{OreshenkoEtal2017apjWASP12bEmissionRetrievals,HimesEtal2022apjWASP12bEmission}.

\subsubsection{3D global circulation model (GCM)}
\label{sect:3D_GCM}
We also compare the measured occultation depths with the output of a forward 3D GCM that self-consistently calculates the thermal emission from the planet. Here, we use \texttt{expeRT/MITgcm} as introduced by \citet{Carone2020, Schneider2022}. The \texttt{expeRT/MITgcm} uses a pre-calculated grid of correlated-k opacities, where we employ here the S1 spectral resolution as described in \citet{Schneider2022} and opacity sources for Na, K, CH$_4$,  H$_2$O, CO$_2$, CO, H$_2$S, HCN, SiO, PH$_3$ and FeH with collision-induced absorption from H$_2$, He and H$^{-}$ and Rayleigh scattering of H$_2$ and He. We omit VO and TiO because our retrievals showed no evidence of a thermal inversion in the upper atmosphere, which would indicate the presence of these molecules.  In post-processing, we calculate the thermal emission of the planet for different orbital phases (including the dayside emission) using a wavelength resolution R=100, where we reduce the spatial resolution of the dayside to 15 degrees in longitude and latitude. During the GCM simulation as well as during post-processing, the abundance of molecules is computed using equilibrium chemistry assuming solar ([Fe/H]=0) metallicity. Thus, we include the effect of thermal ionization of H$_2$ and H$_2$O on our predicted phase curves and dayside emission spectrum.

The dayside emission of the GCM is overplotted in Fig.\,\ref{fig:em_spec} where we see a good agreement with the retrieval model. In particular, the GCM model correctly predicts that water features are muted in the HST/WFC3 wavelength range (between 1.1 and 1.6$\mu$m) due to a combination of thermal ionization and thus dissociation of H$_2$O over large parts of the dayside and electron opacities that suppress molecular features in this wavelength range. For larger wavelengths, where \spitzer (and now JWST) is sensitive, molecular features are present again. Although the GCM dayside emission yields smaller amplitude features compared to the best retrieval model, it is still qualitatively in agreement with the available data.

Using the phase-resolved thermal emission spectrum from the GCM, we integrate the flux with the \cheops and \tess response functions to obtain the thermal phase variation of the planet in both bands. We overplot the thermal GCM along with the fitted planetary models in Fig.\,\ref{fig:PCfit}. For \cheops, the lower amplitude of the thermal GCM compared to the fitted $F_p$ models indicates the need for additional flux from reflection. For \tess, the computed thermal GCM is a good representation of the observed atmospheric phase variation, although the maximum flux and hotspot offset are slightly overestimated; as often seen when comparing GCMs to observations \citep{Parmentier2018, Zhang2018PhaseTemperature}. For ultra-hot Jupiters like WASP-12b, it also has been recognized that their daysides are so hot that the gas is dominated by atoms and ions and no longer by molecules like H$_2$O, TiO, and VO. The locally high gas temperatures, therefore, lead to a moderate ionization of the atmosphere \citep{Arcangeli2018WASP-18b, Parmentier2018FromContext}. While \texttt{expeRT/MITgcm} takes the H$_2$O thermal stability
into account by calculating a chemical equilibrium model and adjusting abundances accordingly, we did not take into account that this partially ionized flow may couple to the exoplanet's magnetic field. Which processes dominate a magnetic coupling will depend on the planet's magnetic field and the local degree of ionization. A partially ionized gas may couple to the magnetic field and some species may, hence, be transported with the flow \citep{2021A&A...648A..80H}. Such interactions may modify the wind flow to dampen the hot spot offset and change the maximum flux \citep{Rogers2014MAGNETICATMOSPHERES,Beltz2022}, in particular in the infrared \citep{May2022}. In this work, we opted to not add drag to the GCM to mimic the unknown magnetic field in \plname and minimize the number of uncertainties.

\subsection{Albedos and dayside/nightside temperatures}
\label{sect:albedos_T}
The observed dayside flux in each passband is composed of the thermal emission from the planet and the reflected light by the atmosphere in that passband. The observed flux is thus given \citep[e.g.,][]{Esteves2013OPTICALEXOPLANETS,Shporer2017-PCreview} as:
\begin{equation}
    \frac{F_d}{F_{\star}} = A_g\left(\frac{R_p}{a}\right)^2 + \frac{\int \mathcal{F}_{\lambda}(T_{\mathrm{p}})\mathcal{T}_{\lambda}d\lambda}{\int \mathcal{F}_{\lambda}(T_{\mathrm{eff}})\mathcal{T}_{\lambda}d\lambda}\left(\frac{R_p}{R_{\star}}\right)^2.
    \label{eqn:depth_contrib}
\end{equation}

\noindent For a planet with radius $R_p$ and semi-major axis $a$, the reflective contribution is determined by the geometric albedo $A_g$, while the thermal contribution is determined by the emission spectra of the planet and the star $\mathcal{F}_{\lambda}(T_{\mathrm{p}})$ and $\mathcal{F}_{\lambda}(T_{\mathrm{eff}})$, respectively. Therefore, interpreting the measured dayside fluxes requires determining the relative contributions of reflection and thermal emission in the \cheops and \tess bands.
 
As seen in the T–P profile in Fig.\,\ref{fig:em_spec}, \cheops and \tess probe similar pressure levels in the atmosphere consistent with black-body dayside temperatures of 2915$\pm$25\,K and 2821$\pm$25\,K, respectively.  The average measured dayside temperature across both bands is $2868\pm17$\,K, in agreement with the effective dayside temperature of $2864\pm15$\,K derived in \citet{Schwartz2017PhaseJupiters}. Using the retrieved thermal emission spectra, we calculated the thermal contribution in the \cheops and \tess passbands as 205$\pm$10 and 480$\pm$19\,ppm, respectively. Subtracting the thermal contribution from the observed occultation depths, we used Eq.\,(\ref{eqn:depth_contrib}) to estimate the geometric albedo in the passbands. We find $A_g$\,=\,0.083$\pm$0.015 in the \cheops band and $A_g$\,=\,0.010$\pm$0.023 in the \tess band, indicating that the atmosphere has non-negligible reflectivity in the \cheops band but much lower reflectivity in the \tess band where $A_g$ is consistent with zero with $2\sigma$ upper limit of 0.06. The derived $A_g$ values are consistent with the results from the shorter wavelengths of \hst/STIS where \citet{Bell2017TheHubble} found $A_g < 0.064$.

The derived low geometric albedo of \plname follows the trend of low reflectivity ($Ag\lesssim0.2$) of UHJs in the optical to near-infrared transition bands \citep{Mallonn2019LowRegime}, which is supported by the difficulty in forming condensates at such high temperatures \citep{Parmentier2018FromContext, Wakeford2017High-temperatureAtmospheres}. Perhaps unlikely, but the higher albedo in the \cheops band may be due to high-temperature condensates (e.g., silicates, Al$_2$O$_3$, CaTiO$_3$) on the western terminator that have been transported from the nightside. Indeed, \citet{Sing2013HSTWASP-12b} found that the best-fit model to the \hst/STIS, \hst/WFC3, and \spitzer transmission spectrum of \plname was Mie scattering by Al$_2$O$_3$ haze but \citet{Bell2017TheHubble} ruled out this scenario based on low dayside reflectivity measured in the \hst/STIS band. They instead favored thermal emission and Rayleigh scattering from atomic hydrogen and helium.

\begin{figure*}[th!]
    \centering
    \includegraphics[width=\linewidth]{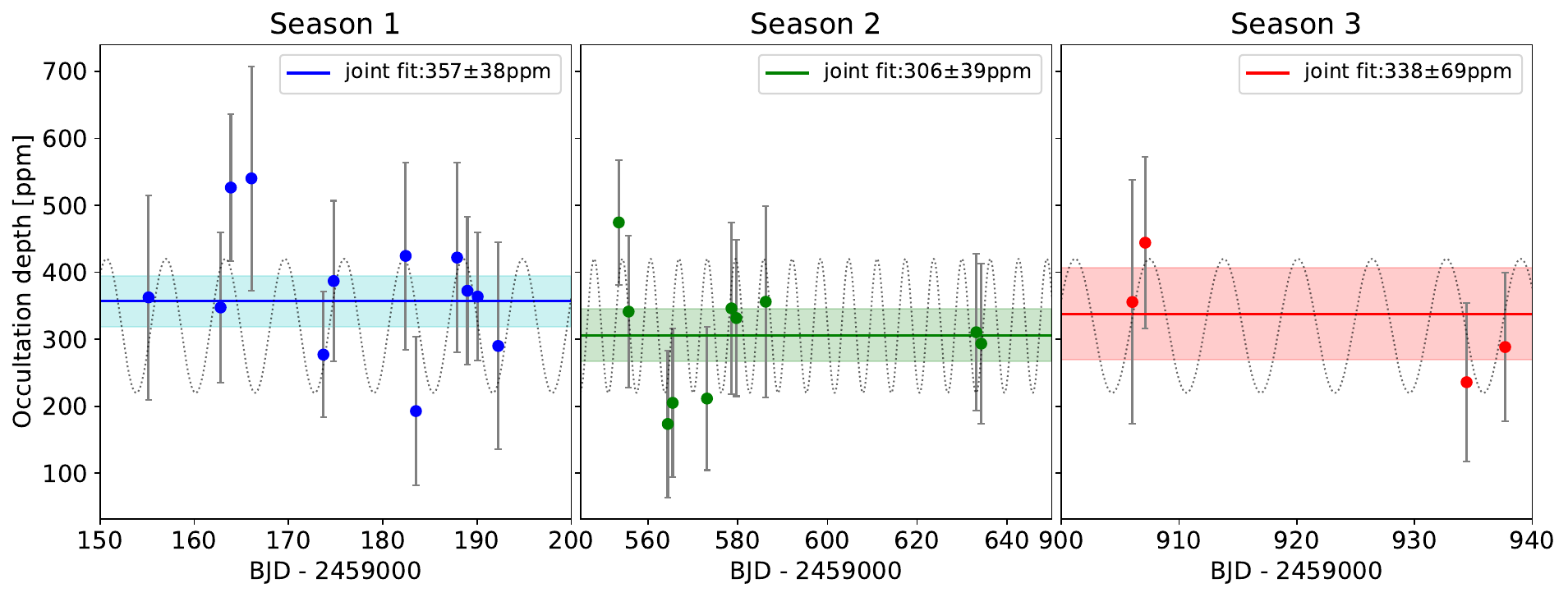}
    \caption{Occultation depth measurements of individual CHEOPS visits in the three observation seasons. The horizontal lines and shaded regions represent the measured depth from jointly fitting all the visits in each season. The depths measured between seasons are consistent within 1$\sigma$. The dotted sinusoidal signal is the best-fit periodic signal (P=8.3\d) to the depth measurements across all seasons.}
    \label{fig:eclipse_var}
\end{figure*}

We used Eq.\,(\ref{eqn:depth_contrib}) to convert the measured nightside fluxes to nightside brightness temperatures and obtained $2\sigma$ upper limits of $T_{\mathrm{night}}\,\simeq\,2000$\,K in both passbands (Table\,\ref{tab:pc_fit}). This implies a large day-night temperature contrast (>45\%) as expected for UHJs due to relatively inefficient heat redistribution \citep{Perna2012THEDISSIPATION,Komacek2017AtmosphericObservations}. The derived high nightside temperature limit is consistent with the expectation of increasing values as a function of stellar irradiation \citep{keating-cowan-dang-2019} as night clouds disperse for highly irradiated planets. 

\subsection{Dayside atmospheric variability}
\label{sect:variability}
Several authors have reported hints of possible time variability in the dayside atmospheric brightness of \plname due to discrepant secondary eclipse depth measurements obtained at various wavelength bands (see Fig.\,\ref{fig:em_spec}). For example, analyses of two i$^\prime$–band observations from different ground-based telescopes resulted in an eclipse depth difference of more than 2$\sigma$ \citep{Hooton2019StormsWASP-12b} while a pair of V–band observations taken within a month of each other revealed eclipse depths that are discrepant with 4.5$\sigma$ significance \citep{VonEssen2019FirstWASP-12b}. There is also a 2.5$\sigma$ discrepancy between published z$^\prime$–band depth measurements \citep{Lopez-Morales2010Day-sideWasp-12b, Fohring2013ULTRACAMWASP-12b} and similar level of disagreement in measurements taken around $\sim$2$\mu$m \cite[see][]{Hooton2019StormsWASP-12b}, and also at \texttt{Spitzer} 3.6$\mu$m and 4.5$\mu$m \citep[e.g.,][]{cowan2012-wasp-12,Stevenson2014TRANSMISSIONm}. Given that the reported eclipse depths have come from different instruments and authors, it is possible that some of the observed disagreements could be due to instrument systematics, observing conditions, telluric contamination in ground-based observations, or even in the analysis of the data. However, \citet{Komacek2019TemporalAtmospheres} showed that the hydrodynamic instabilities in hot Jupiter atmospheres can impact the thermal emission leading to variability at the 2\% level in eclipse depth measurements. \plname can particularly show variability due to magnetohydrodynamic effects in the partially ionized atmosphere \citep{Rogers2014}. It remains unclear whether the observed disagreements are astrophysical or due to systematics. 

Contrary to these results, the analysis of 4 sectors of \tess data by \citet{Wong2022-WASP-12} found no strong evidence of variability between the individual eclipse measurements, although the largest discrepancy between any two measurements is 3.1$\sigma$. Given the low signal-to-noise of the individual \tess eclipses, they were only able to place 2$\sigma$ upper limits of 450\,ppm and 80\,ppm on orbit-to-orbit and month-long variability, respectively.

We further investigate this potential variability with the \cheops occultation observations spanning 3 seasons. The analyses of the 26 occultations have been described in Section\,\ref{sect:occult_analysis}. Figure\,\ref{fig:eclipse_var} shows the derived occultation depths and their $1\sigma$ uncertainties ordered chronologically. The depths derived from the joint fit of occultations with each season of observation are also shown. The individual occultation depths agree with each other within 1$\sigma$ and also with the joint fit of each season better than 1.2$\sigma$. The largest discrepancy between any two depth measurements is 1.8$\sigma$ while the joint fits for the seasons are in agreement within 1$\sigma$. We search for periodicity in the occultation depths using a Lomb-Scargle periodogram \citep{Lomb1976, Scargle1982} and found the highest peak periodicity of 8.3\,d to be nonsignificant with a false alarm probability of $\sim$20\%. Therefore, we do not find strong evidence for variability in the dayside atmosphere of \plname at the median depth precision of 120\,ppm attained by \cheops.

\section{Tidal decay}
\label{sect:tidal_decay}
Previous analyses of the transit times of \plname have clearly shown that the orbit of the planet is decaying due to tidal interaction with the star, causing the planet to lose angular momentum to the star \citep{MacIejewski2016DepartureB, Yee2019TheDecaying, Turner2021DecayingObservations, Wong2022-WASP-12}. The estimate of the planet's decay rate was possible due to the long baseline of available timing measurements. Since our \cheops observations further extend the time baseline, we perform a fit to refine the ephemeris and decay rate of the \plname by combining our transit and occultation timing measurements given in Table\,\ref{tab:transit_times}. We combined these \cheops transit timings with prior transit and occultation timings compiled by \citet{Yee2019TheDecaying} from various authors and \tess timings derived in \citet{Wong2022-WASP-12}.

We model the orbital decay of the planet using a quadratic ephemeris model which gives the transit and occultation times \citep[e.g.,][]{Turner2021DecayingObservations} as:
\begin{equation}
    \begin{aligned}
    \label{eqn:orb_decay}
        T_{\mathrm{tra}} (E) &= T_0 + PE + \frac{1}{2}\frac{dP}{dE}E^2\,, \\
        T_{\mathrm{occ}} (E) &= \left(T_0 + \frac{P}{2} \right) + PE + \frac{1}{2}\frac{dP}{dE}E^2\,
    \end{aligned}
\end{equation}

\noindent where $T_0$ is the reference transit time closest to the middle of the entire time series, $E$ is the transit epoch, and $dP/dE$ is the orbital decay rate from which the time decay rate $\Dot{P} = \frac{1}{P}\frac{dP}{dE}$ and the decay timescale $\tau = P/\dot{P}$ can be derived. We used \texttt{emcee} to simultaneously fit Eq.\,(\ref{eqn:orb_decay}) to the transit and occultation times with $T_0$, $P$, and $dP/dE$ as free parameters. The result from the model fit is given as:

\begin{equation}
    \begin{aligned}
        T_0 &= 2457103.283661\pm3.1\times 10^{-05} \mathrm{BJD_{TBD}},\\
        P &= 1.091419366\pm2\times 10^{-08} \mathrm{days},\\
        \frac{dP}{dE} &= (-1.043\pm0.029)\times 10^{-09} \mathrm{days/orbit},
    \end{aligned}
\end{equation}

\noindent from which we derived:
\begin{equation}
    \begin{aligned}
        \dot{P} &= -30.13\pm0.82\, \mathrm{ms/yr}\\
        \tau &= 3.13\pm0.087\,\mathrm{Myr}.
    \end{aligned}
\end{equation}

\begin{figure}[t!]
    \centering
    \includegraphics[width=\linewidth]{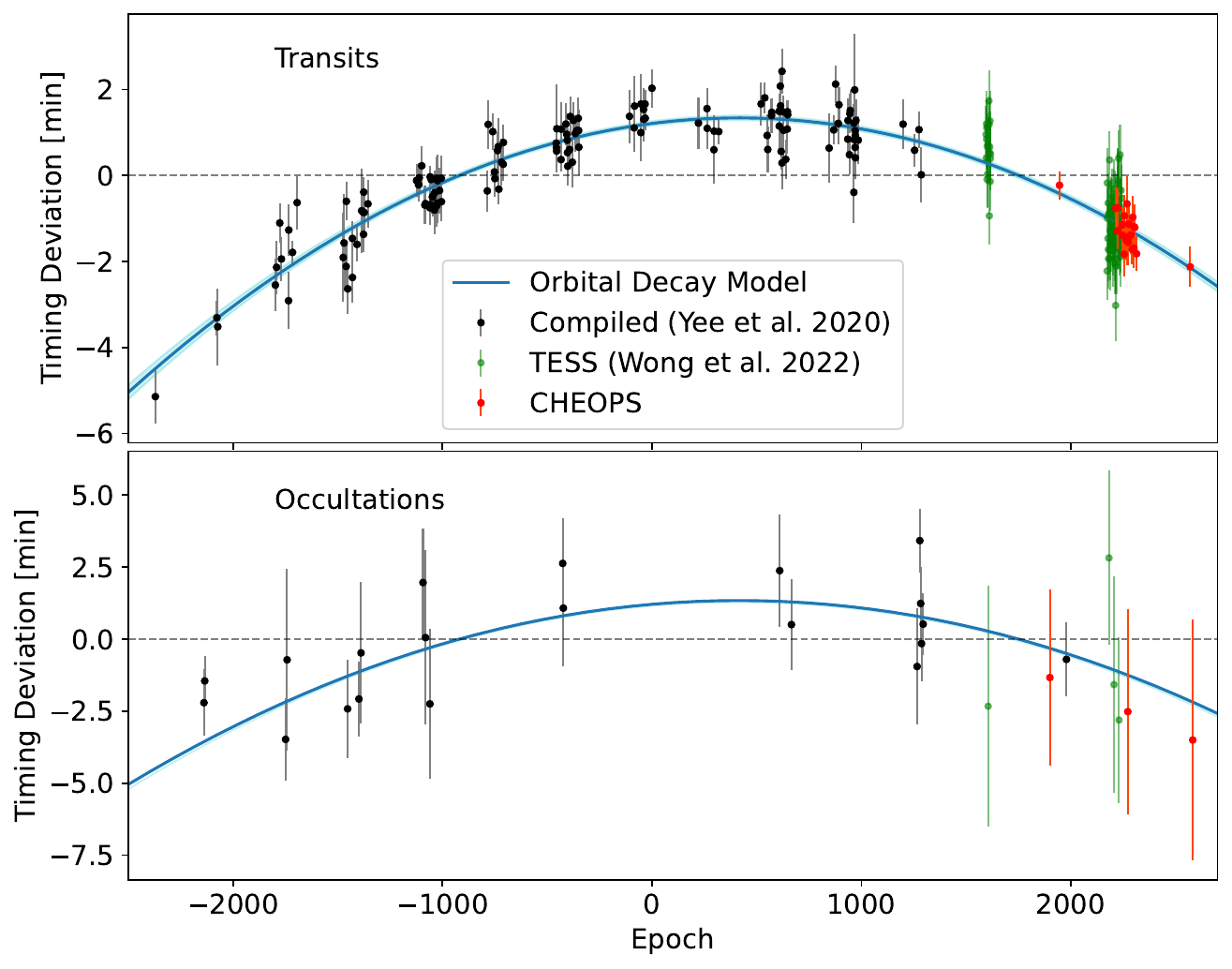}
    \caption{Deviation in the transit (top) and occultation (bottom) timings of WASP-12b compared to the best-fit linear ephemeris model. The black and green points are published measurements taken from \citet{Yee2019TheDecaying} and \citet{Wong2022-WASP-12}, respectively, while the red points are the new \texttt{CHEOPS} timing measurements.}
    \label{fig:orbital_decay}
\end{figure}

The decay rate $\dot{P}$ derived from our fit including new timing measurements by \cheops remains consistent with the estimate of $-29.81\pm0.94$\,ms/yr obtained in \citet{Wong2022-WASP-12} but improves the precision by 12\%. Our revised orbital decay timescale $\tau$ is slightly shorter but 13\% more precise than the value of 3.16\,$\pm$\,0.1\,Myr derived in \citet{Wong2022-WASP-12}. Figure\,\ref{fig:orbital_decay} shows the transit and occultation timing deviations of \plname{} and the orbital decay model fit after subtraction of the best-fit linear ephemeris model.

The rate at which the planet's orbital energy ($E_p$) and angular momentum ($L_p$) are being lost to the star can be calculated \citep[e.g.,][]{Yee2019TheDecaying} as:

\begin{equation}
    \begin{aligned}
    \frac{dE_p}{dt} &= \frac{M_p}{3}\left(\frac{2\pi GM_{\star}}{P}\right)^{2/3}\frac{\dot{P}}{P} = (-5.1\pm0.3)\times 10^{30}\,\mathrm{erg/s}\\
    \frac{dL_p}{dt} &= \frac{M_p\dot{P}}{3(2\pi)^{1/3}}\left(\frac{GM_{\star}}{P}\right)^{2/3} = (-7.64 \pm 0.45 )\times 10^{27}\,\mathrm{kg\,m^2\,s^{-2}}.
    \end{aligned}
\end{equation}

\noindent The energy is then dissipated inside the star as the tidal oscillations are converted into heat. The efficiency of tidal dissipation is quantified by the 
modified tidal quality factor of the star $Q'_{\star}$ which can we derive from $\dot{P}$ using the constant lag model of \citet{Goldreich1966QSystem} as:

\begin{equation}
    Q'_{\star} = -\frac{27\pi}{2}Q_M\left(\frac{a}{R_{\star}}\right)^{-5} \frac{1}{\dot{P}} = (1.70\pm0.14)\times10^{5}.
    \label{eqn:qstar}
\end{equation}

\noindent Population studies of stars show that the value of $Q'_{\star}$ ranges between $10^{5}$–$10^{6.5}$ for hot Jupiter systems and may extend up to $10^7$ for binary star systems \citep{Jackson2008TidalPlanets,Ogilvie2014TidalPlanets}. Lower values of $Q'_{\star}$ imply more efficient tidal dissipation. Our derived value  in agreement with the value of  $Q'_{\star}$\,=\,(1.75$\pm$0.12)$\,\times\,10^{5}$  derived in \citet{Yee2019TheDecaying}. The derived value is on the low end of the range of $Q'_{\star}$ values and implies a much higher dissipation rate for \stname than is expected for a main-sequence star \citep{Turner2021DecayingObservations,Yee2019TheDecaying}. Since the efficiency of tidal dissipation depends also on the structure and evolutionary
state of the star, one explanation would be that \stname is actually a subgiant star capable of such efficient dissipation due to nonlinear breaking of gravity waves close to the center of the star \citep{Weinberg2017TidalWASP-12}. However, further stellar modeling reports that the observed characteristics of \stname are consistent with a main-sequence star rather than a subgiant \citep{Bailey2019UnderstandingWASP-12b}. Our modeling also confirms a young stellar age of 2.3\,Gyr (\S\,\ref{sect:star_pars}).

Tidal decay has not been confirmed for any other exoplanet despite several ultra-hot Jupiter systems having similar planetary and orbital characteristics as \stname. Recently, \citet{Vissapragada2022TheSystem} reported evidence for the tidal decay of Kepler-1658\,b orbiting an evolved star and derived a decay rate of $131_{-22}^{+20}$\,ms/yr. However, further observation of this system will be needed to improve the precision of the orbital decay rate. As this is the first reported case of planetary inspiral around an evolved star, its confirmation will give support to the theoretical expectation of planetary engulfment with stellar evolution, which results in the observed dearth of hot Jupiters around evolved stars \citep{Grunblatt2022TESSStars}. \citet{Harre2023ExaminingB} also reported a $5\sigma$ significant measurement of orbital decay for WASP-4\,b but found that more observations are required to differentiate between the tidal decay and apsidal precession scenarios.

\section{Conclusion}
In this work, we have presented a detailed analysis of \plname using observations by the \cheops spacecraft alongside publicly available data from \tess and \spitzer. We leverage these datasets to put constraints on the shape, atmosphere, and orbital characteristics of the planet. We summarize the main results below:

\begin{itemize}
    \item The large number of \cheops visits allowed us to construct a phase curve which we analyzed together with \tess phase curve and \spitzer transits to constrain the tidal deformation and atmospheric properties of the planet. From our global fit to the datasets, we measured a Love number of \hf{}\,=\,1.55$^{+0.45}_{-0.49}$ corresponding to a 3.16$\sigma$ detection. This measurement makes \plname the second planet, after WASP-103\,b, where tidal deformation has been significantly detected from the light curve. The 3$\sigma$ Love number measurements for these planets are still consistent with those of Jupiter despite the strong irradiation of these planets. There is a need to improve the precision of Love number measurements in order to perform comparative interior structure analysis between these highly irradiated, tidally locked planets and the cooler Jupiter. Phase curve observations of such planets with \texttt{JWST} will provide the precisions needed to measure the Love number at $\sim$12$\sigma$ significance, which will in turn better constrain their core mass fractions and interior structures \citep{Akinsanmi2023OnCurves}.

    \item We significantly measure occultation depths of $333\pm24$\,ppm and $493\pm29$\,ppm in the \cheops and \tess bands, respectively. The nightside fluxes are consistent with zero at both bands. We compared our phase curve with the output of 3D-GCM and found close agreement between them. We measured marginal eastward phase offset of in both bands. We also detected stellar ellipsoidal variation in both passbands, which leads to mass estimates that marginally differ from the RV-derived mass at the 1.2–1.8$\sigma$ level if a spherical planet shape is assumed. However, when we account for tidal deformation, the mass estimates agree with the RV value at $<1\sigma$ indicating a preference for the ellipsoidal model.

    \item We model the emission spectrum of \plname using published occultation depth measurements spanning 0.5-8$\mu$m to derive the thermal profile of the planet. Our best-fit model indicates that \cheops and \tess are probing the same pressure level in the atmosphere of \plname with an average brightness temperature of $\sim$2868\,K.  We found no evidence of temperature inversion in the atmosphere in agreement with the conclusion from \citet{madhu_wasp-12}. We additionally estimate the geometric albedo of the planet and find the planet to be more reflective in the \cheops band with $A_g=0.086\pm0.017$ than in \tess with $A_g=0.01\pm0.023$. 

    \item Our analysis of the \cheops occultations did not show strong evidence of variability in the dayside atmosphere of the planet at the median occultation depth precision of 120\,ppm attained by \cheops. 

    \item Our analysis of the tidal decay of the planet using the new \cheops observations refines the orbital decay rate of the planet to $-30.13\pm0.82$\,ms/yr corresponding to a precision improvement of 12\% compared to the latest estimates from \citet{wong2021_tess2ndyearPCs}. 

\end{itemize}
\label{sect:conclusion}
\begin{acknowledgements}
CHEOPS is an ESA mission in partnership with Switzerland with important contributions to the payload and the ground segment from Austria, Belgium, France, Germany, Hungary, Italy, Portugal, Spain, Sweden, and the United Kingdom. The CHEOPS Consortium would like to gratefully acknowledge the support received by all the agencies, offices, universities, and industries involved. Their flexibility and willingness to explore new approaches were essential to the success of this mission. 
BA acknowledges the financial support of the Swiss National Science Foundation under grant number PCEFP2\_194576. 
ML acknowledges support of the Swiss National Science Foundation under grant number PCEFP2\_194576. 
S.C.C.B. acknowledges support from FCT through FCT contracts nr. IF/01312/2014/CP1215/CT0004. 
LCa and CHe acknowledge support from the European Union H2020-MSCA-ITN-2019 under Grant Agreement no. 860470
(CHAMELEON). 
P.E.C. is funded by the Austrian Science Fund (FWF) Erwin Schroedinger Fellowship, program J4595-N. 
ACCa and TWi acknowledge support from STFC consolidated grant numbers ST/R000824/1 and ST/V000861/1, and UKSA grant number ST/R003203/1. 
GBr, GSc, VSi, LBo, VNa, IPa, GPi, and RRa acknowledge support from CHEOPS ASI-INAF agreement n. 2019-29-HH.0. 
B.-O. D. acknowledges support from the Swiss State Secretariat for Education, Research and Innovation (SERI) under contract number MB22.00046. 
ABr was supported by the SNSA. 
S.G.S. acknowledge support from FCT through FCT contract nr. CEECIND/00826/2018 and POPH/FSE (EC). 
This work was supported by FCT - Fundação para a Ciência e a Tecnologia through national funds and by FEDER through COMPETE2020 - Programa Operacional Competitividade e Internacionalizacão by these grants: UID/FIS/04434/2019, UIDB/04434/2020, UIDP/04434/2020, PTDC/FIS-AST/32113/2017 \& POCI-01-0145-FEDER- 032113, PTDC/FIS-AST/28953/2017 \& POCI-01-0145-FEDER-028953, PTDC/FIS-AST/28987/2017 \& POCI-01-0145-FEDER-028987, O.D.S.D. is supported in the form of work contract (DL 57/2016/CP1364/CT0004) funded by national funds through FCT. 
YAl acknowledges support from the Swiss National Science Foundation (SNSF) under grant 200020\_192038. 
RAl, DBa, EPa, and IRi acknowledge financial support from the Agencia Estatal de Investigación of the Ministerio de Ciencia e Innovación MCIN/AEI/10.13039/501100011033 and the ERDF “A way of making Europe” through projects PID2019-107061GB-C61, PID2019-107061GB-C66, PID2021-125627OB-C31, and PID2021-125627OB-C32, from the Centre of Excellence “Severo Ochoa'' award to the Instituto de Astrofísica de Canarias (CEX2019-000920-S), from the Centre of Excellence “María de Maeztu” award to the Institut de Ciències de l’Espai (CEX2020-001058-M), and from the Generalitat de Catalunya/CERCA programme. 
XB, SC, DG, MF and JL acknowledge their role as ESA-appointed CHEOPS science team members. 
This work has been carried out within the framework of the NCCR PlanetS supported by the Swiss National Science Foundation under grants 51NF40\_182901 and 51NF40\_205606. 
This project was supported by the CNES. 
The Belgian participation to CHEOPS has been supported by the Belgian Federal Science Policy Office (BELSPO) in the framework of the PRODEX Program, and by the University of Liège through an ARC grant for Concerted Research Actions financed by the Wallonia-Brussels Federation. 
L.D. is an F.R.S.-FNRS Postdoctoral Researcher. 
This project has received funding from the European Research Council (ERC) under the European Union’s Horizon 2020 research and innovation programme (project {\sc Four Aces}. 
grant agreement No 724427). It has also been carried out in the frame of the National Centre for Competence in Research PlanetS supported by the Swiss National Science Foundation (SNSF). DE acknowledges financial support from the Swiss National Science Foundation for project 200021\_200726. 
MF and CMP gratefully acknowledge the support of the Swedish National Space Agency (DNR 65/19, 174/18). 
DG gratefully acknowledges financial support from the CRT foundation under Grant No. 2018.2323 ``Gaseousor rocky? Unveiling the nature of small worlds''. 
M.G. is an F.R.S.-FNRS Senior Research Associate. 
MNG is the ESA CHEOPS Project Scientist and Mission Representative, and as such also responsible for the Guest Observers (GO) Programme. MNG does not relay proprietary information between the GO and Guaranteed Time Observation (GTO) Programmes, and does not decide on the definition and target selection of the GTO Programme. 
SH gratefully acknowledges CNES funding through the grant 837319. 
KGI is the ESA CHEOPS Project Scientist and is responsible for the ESA CHEOPS Guest Observers Programme. She does not participate in, or contribute to, the definition of the Guaranteed Time Programme of the CHEOPS mission through which observations described in this paper have been taken, nor to any aspect of target selection for the programme. 
K.W.F.L. was supported by Deutsche Forschungsgemeinschaft grants RA714/14-1 within the DFG Schwerpunkt SPP 1992, Exploring the Diversity of Extrasolar Planets. 
This work was granted access to the HPC resources of MesoPSL financed by the Region Ile de France and the project Equip@Meso (reference ANR-10-EQPX-29-01) of the programme Investissements d'Avenir supervised by the Agence Nationale pour la Recherche. 
AC acknowledges support from PTDC/FIS-AST/7002/2020. 
PM acknowledges support from STFC research grant number ST/M001040/1. 
This work was also partially supported by a grant from the Simons Foundation (PI Queloz, grant number 327127). 
NCSa acknowledges funding by the European Union (ERC, FIERCE,101052347). Views and opinions expressed are however those of the author(s) only and do not necessarily reflect those of the European Union or the European Research Council. Neither the European Union nor the granting authority can be held responsible for them. 
GyMSz acknowledges the support of the Hungarian National Research, Development and Innovation Office (NKFIH) grant K-125015, a PRODEX Experiment Agreement No. 4000137122, the Lend\"ulet LP2018-7/2021 grant of the Hungarian Academy of Science and the support of the city of Szombathely. 
V.V.G. is an F.R.S-FNRS Research Associate. 
NAW acknowledges UKSA grant ST/R004838/1.

\end{acknowledgements}

%
%

\bibliographystyle{aa} 
\bibliography{references, auto_references}

\begin{appendix}
\section{Tables}
\FloatBarrier
\begin{table*}
\caption{\cheops observation log of \stname. The visit types are either occultation (occ), transit (tra), or phase-curve (PC). $\beta_w\beta_r$ gives the white and red noise correction factor to the flux uncertainties of each visit.}
\label{tab:observations}
\resizebox{\textwidth}{!}{%

\begin{tabular}{cccccccccl}
\hline\hline
\begin{tabular}[c]{@{}c@{}}Visit\\ {[}\#{]}\end{tabular} &
  File key &
  Type &
  \begin{tabular}[c]{@{}c@{}}Start time\\ {[}UTC{]}\end{tabular} &
  \begin{tabular}[c]{@{}c@{}}Duration\\ {[}hr{]}\end{tabular} &
  \begin{tabular}[c]{@{}c@{}}Data\\ points {[}\#{]}\end{tabular} &
  \begin{tabular}[c]{@{}c@{}}Eff.\\ {[}\%{]}\end{tabular}  & Epoch & $\beta_w\beta_r$   \\ \hline
1  & PR100016\_TG010201\_V0200 & occ & 2020-11-02T00:45:45  & 7.1  & 223 & 52 & 1880  & 1.54  \\
2  & PR100016\_TG010202\_V0200 & occ & 2020-11-09T16:06:45  & 7.8  & 249 & 52 & 1887  & 1.02  \\
3  & PR100016\_TG010203\_V0200 & occ & 2020-11-10T19:34:19  & 5.8  & 212 & 61 & 1888  & 1.01  \\
4  & PR100016\_TG010204\_V0200 & occ & 2020-11-12T22:41:45  & 6.8  & 226 & 55 & 1890  & 1.13  \\
5  & PR100016\_TG010205\_V0200 & occ & 2020-11-20T14:02:20  & 7.8  & 253 & 53 & 1897  & 1.01  \\
6  & PR100016\_TG010206\_V0200 & occ & 2020-11-21T16:14:20  & 6.8  & 236 & 57 & 1898  & 1.10  \\
7  & PR100016\_TG010207\_V0200 & occ & 2020-11-29T07:35:20  & 8.2  & 244 & 49 & 1905  & 1.10  \\
8  & PR100016\_TG010208\_V0200 & occ & 2020-11-30T10:12:20  & 6.7  & 220 & 54 & 1906  & 1.16  \\
9  & PR100016\_TG010209\_V0200 & occ & 2020-12-04T19:03:20  & 6.8  & 237 & 57 & 1910   & 1.05  \\
10 & PR100016\_TG010210\_V0200 & occ & 2020-12-05T21:16:20  & 5.9  & 223 & 63 & 1911   & 0.96  \\
11 & PR100016\_TG010211\_V0200 & occ & 2020-12-06T23:37:22  & 7.1  & 251 & 58 & 1912   & 1.05  \\
12 & PR100016\_TG010212\_V0200 & occ & 2020-12-09T03:21:20  & 6.8  & 223 & 54 & 1914   & 1.41 \\
13 & PR100013\_TG001201\_V0200 & tra & 2021-01-12T11:10:20  & 8.9  & 331 & 62 & 1946   & 1.43 \\
14 & PR100013\_TG001701\_V0200 & tra & 2021-11-01T00:53:20  & 9.1  & 283 & 51 & 2214  & 1.11 \\
15 & PR100013\_TG001702\_V0200 & tra & 2021-11-10T19:40:22  & 9.8  & 303 & 51 & 2223  & 1.01 \\
16 & PR100013\_TG001703\_V0200 & tra & 2021-11-11T22:59:21  & 9.1  & 298 & 54 & 2224  & 1.19 \\
17 & PR100016\_TG015001\_V0200 & occ & 2021-12-05T09:14:25  & 11.5 & 384 & 55 & 2245  & 1.00 \\
18 & PR100013\_TG001704\_V0200 & tra & 2021-12-05T20:55:21  & 10.5 & 361 & 57 & 2246  & 0.93 \\
19 & PR100013\_TG001705\_V0200 & tra & 2021-12-07T00:38:23  & 10.6 & 362 & 56 & 2247  & 0.93 \\
20 & PR100016\_TG015002\_V0200 & occ & 2021-12-07T11:28:21  & 11.1 & 373 & 55 & 2247  & 1.08 \\
21 & PR100013\_TG001706\_V0200 & tra & 2021-12-15T16:39:21  & 12.2 & 424 & 58 & 2255  & 1.12 \\
22 & PR100016\_TG015003\_V0200 & occ & 2021-12-16T05:36:20  & 12.4 & 441 & 59 & 2255  & 1.33 \\
23 & PR100013\_TG001707\_V0200 & tra & 2021-12-16T18:51:21  & 12.2 & 425 & 58 & 2256  & 1.27 \\
24 & PR100016\_TG015004\_V0200 & occ & 2021-12-17T07:12:21  & 12.6 & 407 & 53 & 2256  & 1.04 \\
25 & PR100016\_TG015005\_V0200 & occ & 2021-12-24T22:55:20  & 10.7 & 379 & 58 & 2263  & 1.21 \\
26 & PR100013\_TG001708\_V0200 & tra & 2021-12-26T14:35:21  & 9.8  & 327 & 55 & 2265  & 1.32 \\
27 & PR100013\_TG001709\_V0200 & tra & 2021-12-29T21:36:21  & 11.7 & 394 & 55 & 2268  & 1.64 \\
28 & PR100016\_TG015006\_V0200 & occ & 2021-12-30T09:32:22  & 13.6 & 479 & 58 & 2268  & 0.98 \\
29 & PR100013\_TG001710\_V0200 & tra & 2021-12-30T23:58:21  & 10.8 & 394 & 60 & 2269  & 1.82 \\
30 & PR100016\_TG015007\_V0200 & occ & 2021-12-31T11:43:21  & 14.9 & 522 & 58 & 2269  & 1.49 \\
31 & PR100013\_TG001711\_V0200 & tra & 2022-01-04T08:15:21  & 9.8  & 331 & 56 & 2273  & 1.68 \\
32 & PR100013\_TG001712\_V0200 & tra & 2022-01-06T12:37:21  & 12.1 & 421 & 58 & 2275  & 1.34 \\
33 & PR100016\_TG015008\_V0200 & occ & 2022-01-07T00:53:22  & 11.7 & 404 & 57 & 2275  & 1.17 \\
34 & PR100013\_TG001713\_V0200 & tra & 2022-01-09T19:41:22  & 12.1 & 429 & 58 & 2278  & 1.10 \\
35 & PR100013\_TG001714\_V0200 & tra & 2022-01-19T15:49:21  & 9.8  & 309 & 52 & 2287  & 1.44 \\
36 & PR100013\_TG001715\_V0200 & tra & 2022-01-26T04:51:21  & 9.8  & 331 & 56 & 2293  & 1.28 \\
37 & PR100013\_TG001716\_V0200 & tra & 2022-01-29T11:12:21  & 9.0  & 314 & 58 & 2296  & 1.35 \\
38 & PR100013\_TG001717\_V0200 & tra & 2022-02-03T23:05:22  & 10.8 & 375 & 58 & 2301  & 1.09 \\
39 & PR100013\_TG001718\_V0200 & tra & 2022-02-04T23:49:21  & 9.1  & 325 & 59 & 2302  & 0.90 \\
40 & PR100013\_TG001719\_V0200 & tra & 2022-02-08T06:57:21  & 10.0 & 328 & 54 & 2305  & 1.08 \\
41 & PR100013\_TG001720\_V0200 & tra & 2022-02-18T03:33:21  & 10.8 & 337 & 51 & 2314  & 1.14 \\
42 & PR100016\_TG015601\_V0200 & occ & 2022-02-23T00:04:21  & 11.4 & 365 & 53 & 2318  & 1.00 \\
43 & PR100016\_TG015602\_V0200 & occ & 2022-02-24T02:07:21  & 10.5 & 352 & 56 & 2319  & 1.22 \\
44 & PR100016\_TG015603\_V0200 & occ & 2022-11-22T19:53:22  & 10.7 & 374 & 58 & 2568  & 1.68 \\
45 & PR100016\_TG015604\_V0200 & PC & 2022-11-23T22:12:21   & 24.0 & 787 & 54 & 2569  & 1.21 \\
46 & PR330093\_TG000201\_V0200 & occ & 2022-12-21T04:47:22  & 8.1  & 270 & 55 & 2594  & 1.00 \\
47 & PR100016\_TG015605\_V0200 & occ & 2022-12-24T11:46:21  & 10.6 & 390 & 61 & 2597  & 1.53 \\ \hline
\end{tabular}%
}
\end{table*}
\FloatBarrier

\begin{table*}
\centering
\caption{\tess observation log of \stname}
\label{tab:tess_obs}
\begin{tabular}{cccccccc}
\hline\hline
Sector & \begin{tabular}[c]{@{}c@{}}Start date\\ {[}UTC{]}\end{tabular}  &\begin{tabular}[c]{@{}c@{}}Duration\\ {[}days{]}\end{tabular} & \begin{tabular}[c]{@{}c@{}}Data\\ points {[}\#{]}\end{tabular} & \begin{tabular}[c]{@{}c@{}}Exp.\\ time {[}s{]}\end{tabular}    & $\beta_w\beta_r$ \\ \hline
20     & 2019-12-24     & 27      & 16552    & 120       & 0.58      \\
43     & 2021-09-16     & 27      & 15577    & 120       & 0.88      \\
44     & 2021-11-12     & 27      & 15777    & 120       & 0.97      \\
45     & 2021-11-06     & 27      & 16085    & 120       & 1.23      \\ \hline
\end{tabular}%
\end{table*}

\begin{table*}
\centering
\caption{\spitzer transit observation log of \stname}
\label{tab:spitzer_obs}
\begin{tabular}{cccccccc}
\hline\hline
\begin{tabular}[c]{@{}c@{}}pass\\band\end{tabular} & \begin{tabular}[c]{@{}c@{}}Start date\\ {[}UTC{]}\end{tabular}  &\begin{tabular}[c]{@{}c@{}}Duration\\ {[}hrs{]}\end{tabular} & \begin{tabular}[c]{@{}c@{}}Data\\ points {[}\#{]}\end{tabular} & \begin{tabular}[c]{@{}c@{}}Exp.\\ time {[}s{]}\end{tabular}    & $\beta_w\beta_r$\\ \hline
3.6$\mu$m     & 2010-11-17      & 9.6   & 264   & 128  & 1.01        \\
4.5$\mu$m     & 2010-12-11      & 9.6   & 266   & 128  & 1.03        \\
3.6$\mu$m     & 2013-12-12      & 9.6   & 264   & 128  & 1.01       \\
4.5$\mu$m     & 2013-12-15      & 9.6   & 266   & 128  & 0.95        \\ \hline
\end{tabular}%
\end{table*}

\begin{table*}[ht]
\centering
\caption{Derived power-2 LDC priors in the different passbands and the posterior from the ellipsoidal and spherical planet model fits.}
\label{tab:ldc_priors}
\begin{tabular}{llll}
\hline
  Parameter &
  \multicolumn{1}{l|}{Prior} &
  \multicolumn{2}{c}{Posterior} \\ \cline{3-4} 
   &
  \multicolumn{1}{l|}{} &
  \begin{tabular}[c]{@{}l@{}}Spherical planet\end{tabular} &
  \begin{tabular}[c]{@{}l@{}}Ellipsoidal planet\end{tabular} \\ \hline

  \begin{tabular}[c]{@{}l@{}}$c_{_{\mathrm{CHEOPS}}}$\\ $\alpha_{_{\mathrm{CHEOPS}}}$\end{tabular} &
  \multicolumn{1}{l|}{\begin{tabular}[c]{@{}l@{}}$\mathcal{N}$(0.714, 0.013)\\ $\mathcal{N}$(0.631, 0.016)\end{tabular}} &
  \begin{tabular}[c]{@{}l@{}}$0.706\pm{0.010}$\\ $0.624\pm{0.010}$\end{tabular} &
  \begin{tabular}[c]{@{}l@{}}$0.714\pm{0.010}$\\ $0.630\pm{0.011}$\end{tabular} \\[0.4cm]
  
  \begin{tabular}[c]{@{}l@{}}$c_{_{\mathrm{TESS}}}$\\ $\alpha_{_{\mathrm{TESS}}}$\end{tabular} &
  \multicolumn{1}{l|}{\begin{tabular}[c]{@{}l@{}}$\mathcal{N}$(0.634,0.018)\\ $\mathcal{N}$(0.554,0.019)\end{tabular}} &
  \begin{tabular}[c]{@{}l@{}}$0.627\pm0.013$\\ $0.542\pm{0.014}$\end{tabular} &
  \begin{tabular}[c]{@{}l@{}}$0.636\pm0.013$\\ $0.548\pm{0.014}$\end{tabular} \\[0.4cm]

  \begin{tabular}[c]{@{}l@{}}$c_{_{\mathrm{Spitzer3.6}}}$\\ $\alpha_{_{\mathrm{Spitzer3.6}}}$\end{tabular} &
  \multicolumn{1}{l|}{\begin{tabular}[c]{@{}l@{}}$\mathcal{N}$(0.313, 0.011)\\ $\mathcal{N}$(0.351, 0.015)\end{tabular}} &
  \begin{tabular}[c]{@{}l@{}}$0.316\pm0.008$\\$0.348\pm0.011$ \end{tabular} &
  \begin{tabular}[c]{@{}l@{}}$0.317\pm0.010$\\$0.346\pm0.011$ \end{tabular} \\[0.4cm]

  \begin{tabular}[c]{@{}l@{}}$c_{_{\mathrm{Spitzer4.5}}}$\\ $\alpha_{_{\mathrm{Spitzer4.5}}}$\end{tabular} &
  \multicolumn{1}{l|}{\begin{tabular}[c]{@{}l@{}}$\mathcal{N}$(0.245, 0.011)\\ $\mathcal{N}$(0.398, 0.022)\end{tabular}} &
  \begin{tabular}[c]{@{}l@{}}$0.241\pm0.010$\\$0.406\pm0.020$ \end{tabular} &
  \begin{tabular}[c]{@{}l@{}}$0.240\pm0.096$\\$0.406\pm0.019$ \end{tabular} \\[0.4cm]
  \hline
    \end{tabular}\\
\end{table*}

\begin{table*}
\centering
\caption{Derived mid-transit times for the individual \cheops transit observations of \plname.}
\label{tab:transit_times}
\begin{tabular}{lllll}
\hline\hline
\multicolumn{1}{c}{\begin{tabular}[c]{@{}c@{}}Visit\\ {[}\#{]}\end{tabular}} & Type & \multicolumn{1}{c}{Epoch} & \multicolumn{1}{c}{\begin{tabular}[c]{@{}c@{}}Timing\\  {[}BJD$_{\mathrm{TBD}}$\,–\,2459000{]}\end{tabular}} & $\sigma_{T}$ [d]\\ \hline
13  &  tra  & 1946  &   227.183907       & 0.000225 \\
14  &  tra  & 2214 &   519.683809       & 0.000351 \\
15  &  tra  & 2223 &   529.506215       & 0.000294 \\
16  &  tra  & 2224 &   530.598008       & 0.000305 \\
18  &  tra  & 2246 &   554.608948       & 0.000286 \\
19  &  tra  & 2247 &   555.700197       & 0.000309 \\
21  &  tra  & 2255 &   564.431242       & 0.000353 \\
23  &  tra  & 2256 &   565.523282       & 0.000336 \\
26  &  tra  & 2265 &   575.345658       & 0.000318 \\
27  &  tra  & 2268 &   578.620503       & 0.000449 \\
29  &  tra  & 2269 &   579.711388       & 0.000458 \\
31  &  tra  & 2273 &   584.076978       & 0.000373 \\
32  &  tra  & 2275 &   586.259887       & 0.000212 \\
34  &  tra  & 2278 &   589.534364       & 0.000313 \\
35  &  tra  & 2287 &   599.356962       & 0.000307 \\
36  &  tra  & 2293 &   605.905245       & 0.00024  \\
37  &  tra  & 2296 &   609.180016       & 0.000348 \\
38  &  tra  & 2301 &   614.636613       & 0.000316 \\
39  &  tra  & 2302 &   615.728374       & 0.000211 \\
40  &  tra  & 2305 &   619.002623       & 0.000373 \\
41  &  tra  & 2314 &   628.824968       & 0.000283 \\
45b &  tra  & 2570 &   908.228003       & 0.000327 \\ 
\hline
S1  & occ  & 1900  &   177.523576       &  0.002121       \\
S2  & occ  & 2272  &   583.530597       & 0.002471       \\
S3  & occ  & 2582  &   921.869782       & 0.002904       \\                         
\hline
\end{tabular}
\end{table*}
\FloatBarrier
\section{Figures}
\FloatBarrier

\begin{figure*}
    \centering
    \includegraphics[width=0.96\textwidth]{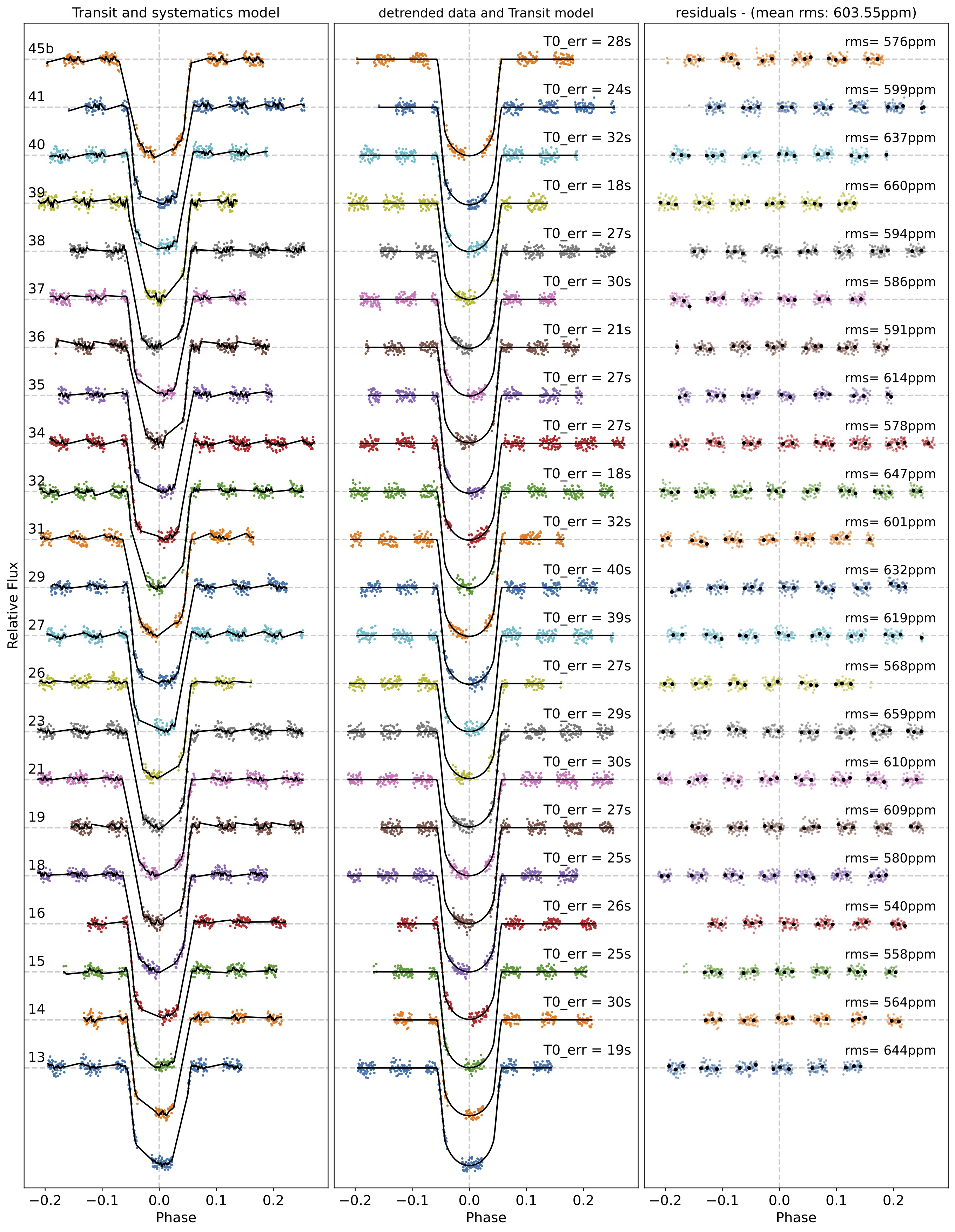}
    \caption{CHEOPS transit light curves of WASP-12b labeled according to the visit number. \textit{Left:} The best-fit transit and systematics model is overplotted on the data. \textit{Middle:} Systematics detrended flux with transit model overplotted. The obtained transit time uncertainty for each visit is shown in seconds. \textit{Right: } Residuals after subtraction of best-fit transit and systematics model. The 30-min bins and the root-mean-square (rms) of each visit are also shown.}
    \label{fig:transit_obs}
\end{figure*}

\begin{figure*}[!ht]
    \centering
    \includegraphics[width=0.96\textwidth]{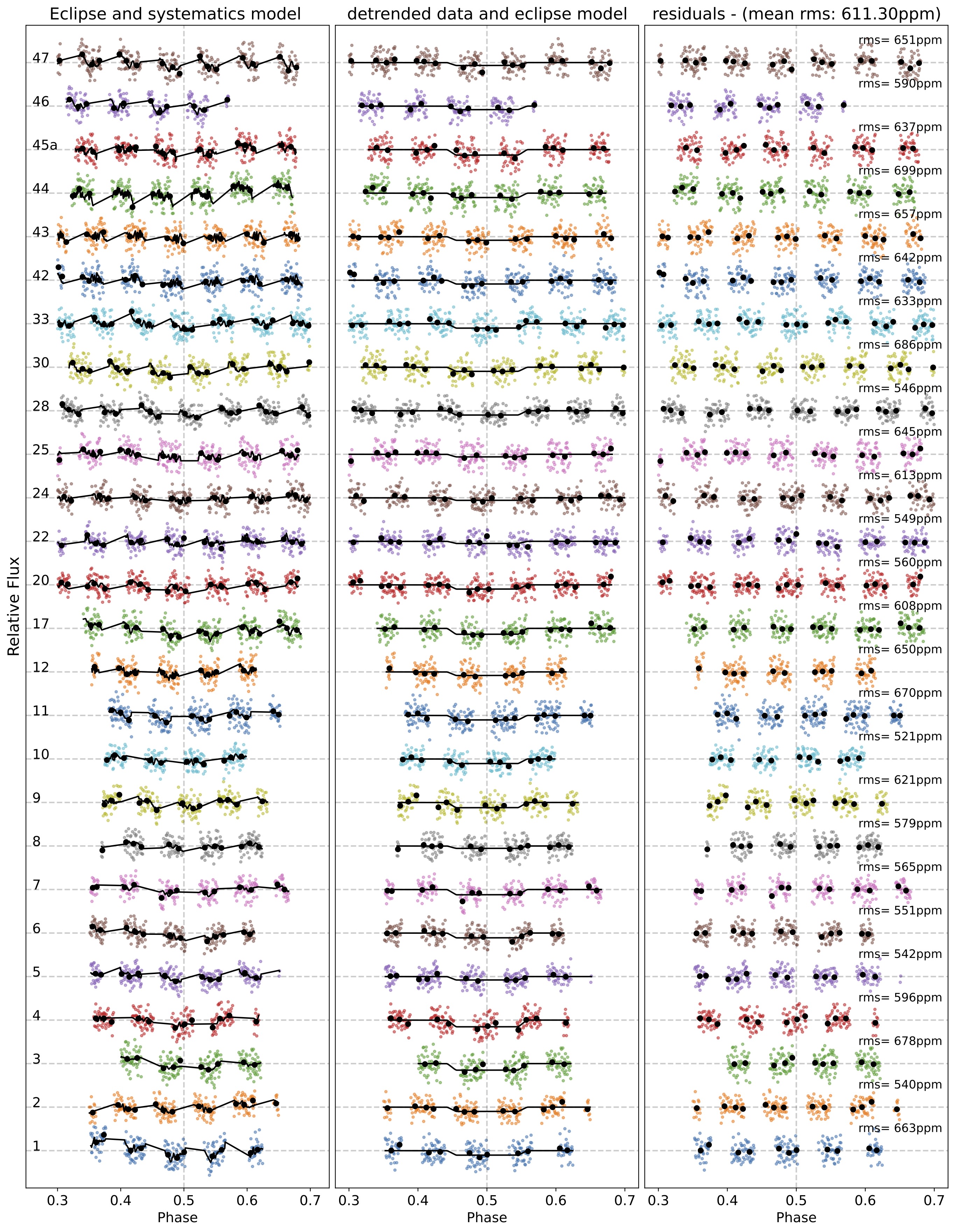}
    \caption{CHEOPS occultation light curves of WASP-12b labeled according to their visit numbers. \textit{Left:} The best-fit occultation and systematics model is overplotted on the data. \textit{Middle:} Systematics detrended flux with occultation model overplotted. \textit{Right:} Residuals after subtraction of best-fit occultation and systematics model. The 30-min bins and the root-mean-square (rms) of each visit are also shown.}
    \label{fig:occulation_obs}
\end{figure*}

\begin{figure*}[!ht]
    \centering
    \includegraphics[width=\linewidth]{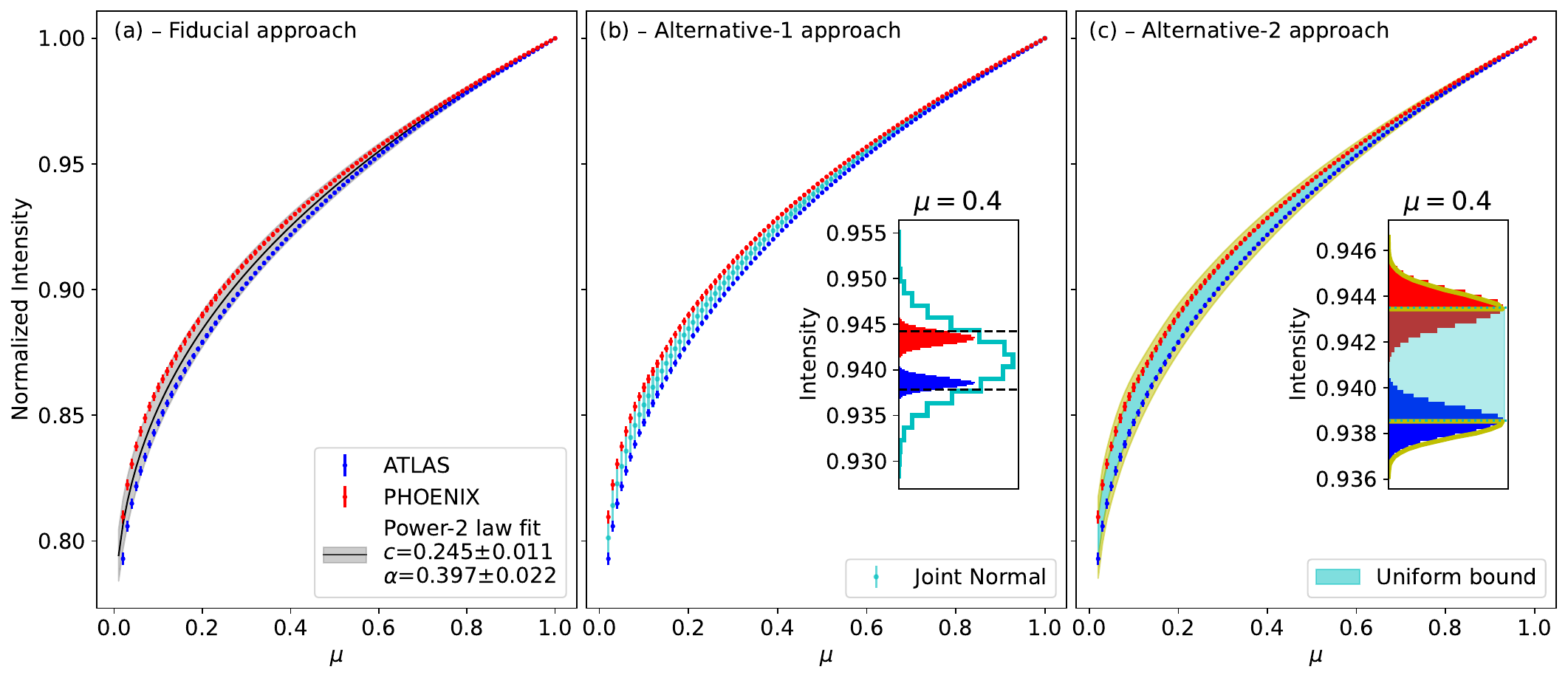}
    \caption{ Different approaches adopted to model the stellar limb darkening using model stellar intensity profiles from the \textsc{PHOENIX} (red) and ATLAS (blue) libraries. The plots illustrate the approaches for the \spitzer 4.5$\mu$m passband observations. Panel (a) illustrates our "fiducial" approach where LDCs are generated to be used as priors in the transit model fitting. The black curve shows the fit of the power-2 law to the combined model intensity profiles. The parameter space (gray) allowed by the 1$\sigma$ uncertainty of the obtained LDCs encompasses both intensity profiles and associated 1$\sigma$ uncertainties. Panel (b) shows the "alternative-1" approach which merges the \textsc{PHOENIX} and ATLAS model profiles to create a new joint intensity profile (cyan) whose 1$\sigma$ uncertainties at each $\mu$ encompasses the 1$\sigma$ uncertainty of the individual profiles. This is illustrated in the inset for $\mu=0.4$. The new joint profile is fitted with an LD law alongside the transit observation at each passband. Panel (c) shows the "alternative-2" approach which similarly combines both intensity profiles but creates a uniform bound (cyan) spanning the median of both profiles such that LD profile points within the bound have equal likelihood but decrease as the profile points deviate from the bound.}
    \label{fig:ldc_plot}
\end{figure*}

\begin{figure*}[h!]
    \centering
    \includegraphics[width=0.56\linewidth]{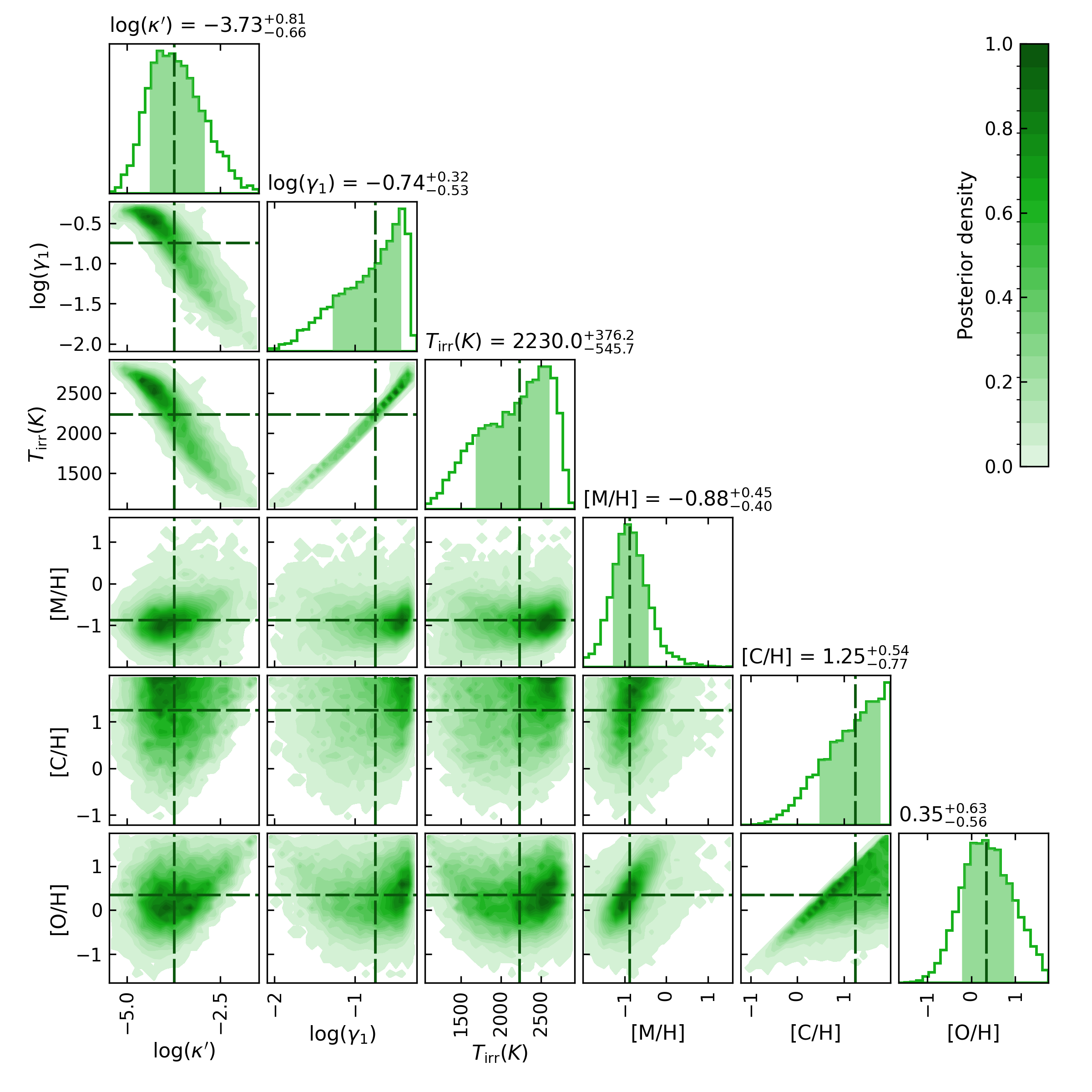}
    \caption{Posterior distribution of the retrieval using {\pyratbay} with the median and $1\sigma$.    }
    \label{fig:pyrat_posterior}
\end{figure*}

\begin{sidewaysfigure*}
    \centering
    \includegraphics[width=0.98\linewidth]{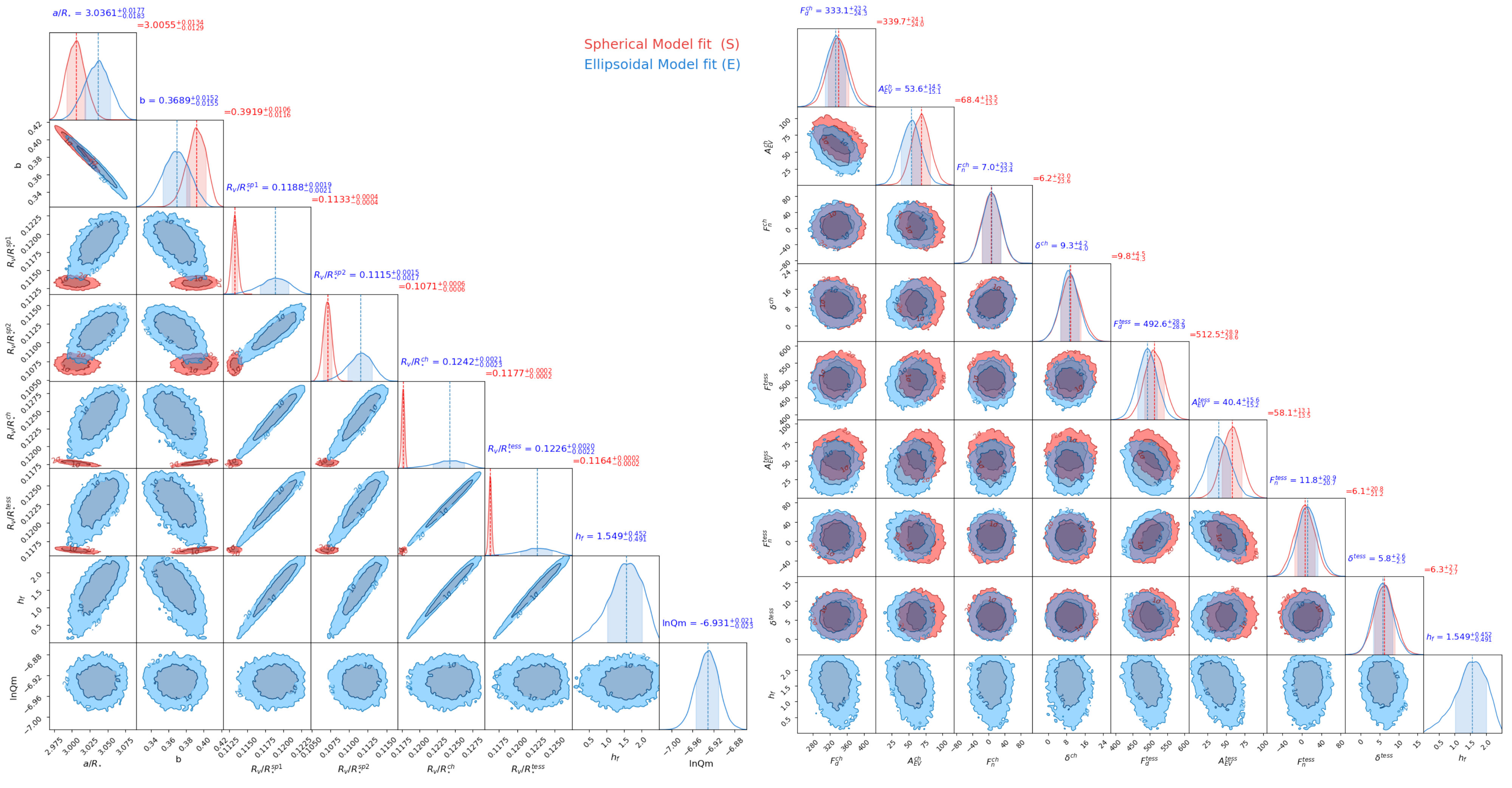}
    \caption{ Corner plot showing the posterior probability distributions of the fitted parameters for the spherical planet model (red) and ellipsoidal planet model. Vertical colored lines indicate the median of the distributions while shaded regions define the 68\%credible intervals. The different instruments are labeled: \cheops (ch), TESS (tess), \spitzer3.6$\mu$m (sp1), and \spitzer4.5$\mu$m (sp2).}
    \label{fig:def_corner}
\end{sidewaysfigure*}
\end{appendix}
\end{document}